\documentclass[aps,prx,twocolumn,nofootinbib,superscriptaddress,reprint,floatfix]{revtex4-2}
\usepackage{amsmath}
\usepackage{bbm}
\usepackage{amssymb}
\usepackage{ifpdf}
\usepackage{color}
\usepackage{xcolor}
\usepackage{graphicx,epsfig}
\usepackage{hyperref}
\usepackage{enumitem}

\hypersetup{
    colorlinks = True,
    linkcolor = {cyan},
    linkbordercolor = {cyan},
    citecolor = {blue},
    citebordercolor = {blue},
    urlcolor = {blue},
}

\usepackage[utf8]{inputenc}
\usepackage{tikz}
\usepackage{pgfplots} 
\usepackage{graphicx}
\usepackage{soul}

\usepackage[normalem]{ulem}

\pgfplotsset{compat=1.16} 
\begin{document}

\title{Dynamical phase transitions in $XY$ model: a Monte Carlo and mean-field theory study}

\author{Mainak Pal}
\affiliation{School of Physical Sciences, Indian Association for the Cultivation of Science, Kolkata 700032, India}

\author{William D. Baez}
\affiliation{Department of Mathematical Sciences, United States Air Force Academy,
Colorado Springs 80840, USA}
\affiliation{Department of Chemistry and Physics, Augusta State University, 2500 Walton Way, Augusta, Georgia 30904, USA}

\author{Pushan Majumdar}
\email[Deceased]{}
\affiliation{School of Physical Sciences, Indian Association for the Cultivation of Science, Kolkata 700032, India}

\author{Arnab Sen}
\email[Corresponding author: ]{tpars@iacs.res.in}
\affiliation{School of Physical Sciences, Indian Association for the Cultivation of Science, Kolkata 700032, India}

\author{Trinanjan Datta}
\email[Corresponding author: ]{tdatta@augusta.edu}
\affiliation{Department of Chemistry and Physics, Augusta State University, 2500 Walton Way, Augusta, Georgia 30904, USA}
\affiliation{Department of Physics and Biophysics, Augusta University, 1120 15th Street, Augusta, Georgia 30912, USA}
\affiliation{Kavli Institute for Theoretical Physics, University of California, Santa Barbara, California 93106, USA}

 \begin{abstract}
   We investigate the dynamical phases and phase transitions arising in a classical two-dimensional anisotropic $XY$ model under the influence of a periodically driven temporal external magnetic field in the form of a symmetric square wave. We use a combination of finite temperature classical Monte Carlo simulation, implemented within a CPU + GPU paradigm, utilizing local dynamics provided by the Glauber algorithm and a phenomenological equation-of-motion approach based on relaxational dynamics governed by the time-dependent free energy within a mean-field approximation to study the model. We investigate several parameter regimes of the variables (magnetic field, anisotropy, and the external drive frequency) that influence the anisotropic $XY$ system. We identify four possible dynamical phases -- Ising-SBO, Ising-SRO, $XY$-SBO and $XY$-SRO. Both techniques indicate that only three of them (Ising-SRO, Ising-SBO, and $XY$-SRO) are stable dynamical phases in the thermodynamic sense. Within the Monte Carlo framework, a finite size scaling analysis,  shows that $XY$-SBO does not survive in the thermodynamic limit giving way to either an Ising-SBO or a $XY$-SRO regime. The finite size scaling  analysis further shows that the transitions between the three remaining  dynamical phases either belong to the two-dimensional Ising universality class or are first-order in nature. Within the mean-field calculations yield three stable dynamical phases, i.e., Ising-SRO, Ising-SBO and $XY$-SRO, where the final steady state is independent of the initial condition chosen to evolve the equations of motion,as well as a region of bi-stability where the system either flows to Ising-SBO or $XY$-SRO (Ising-SRO) depending on the initial condition. Unlike the stable dynamical phases, the $XY$-SBO represents a transient feature that is eventually lost to either Ising-SBO or $XY$-SRO. Our mean-field analysis highlights the importance of the competition between switching of the stationary point(s) of the free energy after each half cycle of the external field and the two-dimensional nature of the phase space for the equations of motion.
 \end{abstract}

\maketitle

\section{\label{sec:intro}Introduction}
The study of the kinetic Ising model~\cite{StollPhysRevB.8.3266,AndersenPhysRevLett.53.1244} and dynamical phase transition (DPT) has a long history. One of the earliest studies of DPT was by Katz, Lebowitz and Spohn who studied non-equilibrium steady states of a stochastic lattice gas under the influence of a static field \cite{Katz1984}. Soon after, the nature of the dynamical response of Ising spins evolving via the Glauber stochastic processes and driven externally by a time periodic magnetic field was investigated by Tom\'{e} and Oliviera \cite{TomePhysRevA.41.4251} using mean-field (MF) techniques. They found that depending on the drive and the bath parameters two distinct types of non-equilibrium steady states, which are related by spontaneous symmetry breaking, can be observed. The system can either oscillate in-phase with the external field or be out-of-phase. The former is referred to as symmetry-restoring oscillations (SRO). The latter is termed symmetry-breaking oscillations (SBO). 

The presence of distinct non-equilibrium steady states and transitions between them were consequently supported by Monte Carlo simulations and large-$N$ ($N$ being the number of components of the spin) expansion calculations \cite{LoPhysRevA.42.7471,PhysRevB.42.856}. 
Extensive theoretical studies focusing on various aspects of DPTs such as universal features of the transition ~\cite{AcharyyaPhysRevE.56.1234,AcharyyaPhysRevE.58.174,KornissPhysRevE.63.016120,KornissPhysRevE.66.056127,AcharyyaPhysRevE.59.218,RikvoldPhysRevE.76.021124,RikvoldPhysRevE.78.051108,SidesPhysRevLett.81.834,PleimlingPhysRevE.87.032145,PleimlingPhysRevLett.109.175703,AKINCI20125810,DevirenPhysRevE.82.022104,KeskinPhysRevE.86.051110}, hysteresis loop scaling behavior~\cite{ZangwillPhysRevE.50.224,AcharyyaPhysRevE.56.2407,AcharyyaPhysRevE.56.1234,AcharyyaPhysRevE.58.179,RikvoldPhysRevE.76.021124} , fluctuation-dissipation relation~\cite{RikvoldPhysRevE.76.021124}, effect of disorder in external field, both spatial \cite{SethnaPhysRevLett.71.3222,SethnaPhysRevLett.70.3347,Basak2020} and temporal \cite{AcharyyaPhysRevE.58.174,AcharyyaXY}, dependence on thermal noise \cite{RikvoldPhysRevLett.92.015701}, and effect of nearest-neighbor and next-nearest neighbor interactions \cite{PhysRevB.31.5946,BAEZ201015} have been performed. For a review on early theoretical development see \cite{AcharyyaRevModPhys.71.847}. In addition to the above computational and theoretical studies, experiments conducted on thin film magnetic materials such as Fe/Au(001) \cite{WangPhysRevLett.70.2336}, thin $p(1\times1)$ Fe films \cite{ErskinePhysRevLett.78.3567}, Cu(001) films \cite{JiangPhysRevB.52.14911,ErskinePhysRevB.59.4249}, Fe/GaAs(001) films \cite{LeePhysRevB.60.10216}, $\text{Ni}_{80}\text{Fe}_{20}$ films \cite{BlandPhysRevB.60.11906}, and Co films~\cite{BergerPhysRevLett.131.116701,BergerPhysRevLett.118.117202,BergerPhysRevLett.111.190602} provide a promising platform to realize and investigate DPT phenomena.

Akin to equilibrium phase transitions, the nearest-neighbor kinetic Ising model in two-dimension ($2d$) has played an important role in understanding DPTs~\cite{SidesPhysRevLett.81.834}.
Monte Carlo (MC) studies of the nearest-neighbor ferromagnetic kinetic Ising (NNKI) model indicates that the system exhibits both the SRO and the SBO phases, where the magnetization oscillates around a zero value in the former and around a non-zero value in the latter dynamical phase. Based on such a perspective, the classical $2d$ ferromagnetic NNKI model provides an accurate description of DPT via magnetization reversal through nucleation and domain wall motion in uniaxial ferromagnets~\cite{KornissPhysRevE.63.016120,KornissPhysRevE.66.056127}. While the NNKI model is an excellent prototype for studying DPT and the ensuing dynamical phases, it does not account for magnetic relaxation processes where a coherent rotation of spins is involved. The study of such a phenomenon requires the spins to rotate through all possible orientations. This fact and the experimental relevance to thin films \cite{BergerPhysRevLett.131.116701,BergerPhysRevLett.118.117202,BergerPhysRevLett.111.190602,JangPhysRevE.63.066119,JiangPhysRevB.52.14911} provides a natural motivation to study DPT in a classical spin model possessing a continuous degree of freedom such as in the classical $2d-XY$ model \cite{Kosterlitz_1973,Kosterlitz_1974}. 

The classical $2d-XY$ model is known to support topological excitations, also known as vortices. In spite of the Mermin-Wagner-Hohenberg theorem~\cite{HohenbergPhysRev.158.383,MerminWagnerPhysRevLett.17.1133}, which prohibits spontaneously broken continuous symmetries at finite temperature in systems with sufficiently short-range interactions in $d\leq 2$, a phase transition related to the binding-unbinding of vortex-antivortex pairs occur at the Berezenskii-Kosterlitz-Thouless (BKT) transition temperature~\cite{Kosterlitz_1973,Kosterlitz_1974}. However, a magnetic field applied along the plane modifies the vortex-antivortex interaction~\cite{FertigPhysRevLett.89.035703,Fertig2003190}. It has been shown that the classical $2d-XY$ magnet in a magnetic field has three distinct vortex phases $-$ a linearly confined phase, a logarithmically confined phase, and a free vortex phase~\cite{Fertig2003190,Fertig2006cond.mat..3028Z}. Additionally, a renormalization group analysis combined with duality in the model shows that at high temperature and high field, the vortices unbind as the magnetic field is lowered in a two-step process. First, strings of overturned spins proliferate. Second, vortices unbind. The transitions are continuous but are not of the Kosterlitz–Thouless type~\cite{FertigPhysRevLett.89.035703}. Thus, it may be natural to ask what happens to the excitations of the equilibrium model when it is exposed to an oscillating magnetic field. 

Yasui~\emph{et al.}~\cite{YasuiPhysRevE.66.036123} have investigated the time-dependent generalization of the $2d-XY$ model in a static magnetic field, both theoretically and numerically, utilizing the anisotropic-$XY$ (an-$XY$) model. The model was analyzed for DPTs using a time-dependent Ginzburg-Landau (continuum) approach. The authors found multiple DPTs which span several dynamical phases: Ising-SRO, $XY$-SRO, Ising-SBO, and $XY$-SBO as summarized in Table~\ref{tab:table1}. The continuous spin system's order parameter exhibits a SRO phase when the frequency of a periodic external field of sufficiently large amplitude is below a critical frequency $\Omega_c(T<T_c,h)$, where $T_c$ refers to the critical temperature of the \textit{zero} field an-$XY$ model (see Eq.~\eqref{eq:hamXY} later in the text for the specification of the model) and $h$ is the external field. When the frequency of the external field is above $\Omega_c(T<T_c,h)$ a SBO phase is observed. The transition between SRO and SBO is an example of DPT, similar to the behavior observed in the NNKI model \cite{SidesPhysRevLett.81.834}. Furthermore, it was shown that all the DPT transitions are in the same universality class as Ising spins in thermal equilibrium in $2d$~\cite{KornissPhysRevE.63.016120, KornissPhysRevE.66.056127}. 
\begin{table}[h]
\caption{\label{tab:table1}
Dynamic order parameter classification of the various phases which arise in the $2d$ an-$XY$ model (see Eq.~\eqref{eq:hamXY}) subject to a spatially uniform and temporally periodic magnetic field. The instantaneous magnetization components along $x$ and $y$ are given by $m_x(t)$ and $m_y(t)$, respectively. The integral is performed over one full cycle where $\Omega$ represents the square wave drive protocol frequency. The dynamic order parameters are later utilized to identify the various dynamical phases of the model. For the Ising model with one-component spin, only one of these components, say the $Q_x$, is relevant.}.
\begin{ruledtabular}
\begin{tabular}{ccc}
  Dynamical Phase &$\displaystyle Q_x$  &$\displaystyle Q_y$  \\ \hline\hline \\
         Ising-SBO & $\displaystyle \frac{\Omega}{2\pi}\oint m_x(t)dt \neq 0$ &$\displaystyle\frac{\Omega}{2\pi}\oint m_y(t)\;dt = 0$ \\ & &  \\
          $XY$-SBO & $\displaystyle\frac{\Omega}{2\pi}\oint m_x(t)\;dt \neq 0$ &$\displaystyle\frac{\Omega}{2\pi}\oint m_y(t)\;dt \neq 0$ \\ & &  \\
           $XY$-SRO & $\displaystyle\frac{\Omega}{2\pi}\oint m_x(t)\;dt = 0$ &$\displaystyle\frac{\Omega}{2\pi}\oint m_y(t)\;dt \neq 0$ \\ & & \\ 
        Ising-SRO & $\displaystyle\frac{\Omega}{2\pi}\oint m_x(t)\;dt = 0$ &$\displaystyle\frac{\Omega}{2\pi}\oint m_y(t)\;dt = 0$  
\end{tabular}
\end{ruledtabular}
\end{table}

The use of Landau formulation with time-averaged coefficients can cause the characteristic features of DPTs to be lost. Moreover, such an analysis can overestimate the stability of one or more dynamical phases. Thus, an unbiased method is needed to clarify the nature of the dynamical phases and DPTs in the 2d an-$XY$ model. We achieve this goal by utilizing the numerical approach of MC simulations with a local Glauber dynamics at finite temperature, implemented within a CPU+GPU paradigm. We investigate the dynamical phases and their corresponding phase transitions that arise in the temporally driven 2d an-$XY$ model. We further develop a MF approach to model this stochastic dynamics by generalizing the approach of Ref.~\onlinecite{TomePhysRevA.41.4251} for Ising spins to the case of two-component $XY$ spins which leads to phenomenological MF equation-of-motion (EOM) in a 2d phase space. In the limit of infinite anisotropy, we recover the standard Ising MF equations of \cite{TomePhysRevA.41.4251} from these generalized MF EOMs. 

Both the MC and the MF techniques were used to investigate several parameter regimes governed by magnetic field, anisotropy, and the external drive frequency. Results from both the approaches suggest that while three of the dynamical phases (Ising-SRO, Ising-SBO, and $XY$-SRO) are stable in the thermodynamic limit, the stability of the $XY$-SBO is more subtle. Within the MC framework, finite size scaling (FSS) analysis indicates that the $XY$-SBO phase does not survive in the thermodynamic limit. Instead, we find either the Ising-SBO or the $XY$-SRO dynamical phases. The FSS analysis further shows that the transitions between the three different dynamical phases either belong to the 2d Ising universality class or are first-order in nature. 

From the perspective of the MF calculations, Ising-SRO, Ising-SBO, and $XY$-SRO represent stable dynamical phases. This is because different initial conditions in the 2d phase space lead to the same final periodic-in-time steady state with {\it{identical}} values for the magnitudes of the corresponding dynamical order parameters. But, the $XY$-SBO phase represents a transient dynamical feature that is eventually lost to either the Ising-SBO or the $XY$-SRO at high frequencies for different sets of initial conditions. In fact, the solutions of the MF EOMs yield Ising-SRO, Ising-SBO, $XY$-SRO and a further region of bistability where different sets of trajectories flow either to the $XY$-SRO or to the Ising-SBO, or either to the Ising-SBO or to the Ising-SRO, respectively.  For small anisotropy and magnetic field, the MF EOMs further predict a prethermal $XY$-SBO phase that eventually relaxes to the $XY$-SRO phase. Thus, the results generated by the two approaches are consistent with each other at a qualitative level. At a conceptual level, the generalized MF EOMs for the $XY$ spins show how trajectories in the 2d phase space effectively concentrate along a 1d line as a function of time for Ising-SBO, Ising-SRO and in the bistable region between the Ising-SBO and the Ising-SRO, while this is not true for the $XY$-SRO phase as well as for the bistable region between Ising-SBO and $XY$-SRO. 

 The organization of the article is as follows. In Sec.~\ref{sec:modmethod} we describe the model and the methods (Monte Carlo simulations and phenomenological mean-field dynamics) used to investigate the 2d an-$XY$ model. In Sec.~\ref{sec:results} we show our results for the an-$XY$ model, provide a discussion on the Monte Carlo and mean-field results, and compare the results with those obtained in the literature using a time dependent Landau-Ginzburg approach. This is followed by an in-depth discussion of the nature of solutions obtained from the MF EOMs, as well as interesting intermediate time dynamics in the region of bistability and for small anisotropy and magnetic field in the same section. In Sec.~\ref{sec:conc} we discuss and provide concluding remarks. More details regarding the MF equations is provided in Appendix~\ref{appendix-MFT}. 
 
\section{\label{sec:modmethod}Model and Methods}
In this section we define 2d an-$XY$ model, describe the MC simulation method based on local Glauber dynamics at finite temperature and the phenomenological MF EOM approach used to study the dynamical phases and DPTs in this model. 
\subsection{Model\label{subsec:model}}
We study the ferromagnetic $2d$ classical an-$XY$ model on the square lattice. The $XY$ spins have uniaxial easy-axis anisotropy $\gamma_a$ along the $x-$direction and a spatially uniform external magnetic field $h^{\text{ext}}_x(t)$ along the $x$-direction which is periodic in time. The Hamiltonian for this periodically driven an-$XY$ model is given by
\begin{equation}
\label{eq:hamXY}
\mathcal{H} = -J\sum_{\langle ij \rangle} \Vec{S}_i \cdot \Vec{S}_j - \gamma_a \sum_i \big(S_{i}^x\big)^2 - h_x^{\text{ext}}(t) \sum_i S_i^{x},
\end{equation}
where the (classical) spin variables $\vec{S}_i$ live on the $2d$ square lattice and can point in any direction in the $2d$ plane. These spins have a first nearest-neighbour ferromagnetic exchange coupling $J>0$. This serves as a natural unit of energy which is  to unity for the remainder of this article. We impose periodic boundary conditions in both directions of the $2d$ square lattice. The $i^{\text{th}}$ spin $\vec{S}_i=(\cos\theta_i,\sin\theta_i)$ is parametrized by the angle $\theta_i$ which it makes with the $x$-axis. Based on this choice Eq.~\eqref{eq:hamXY} can be rewritten as
\begin{equation}
\mathcal{H} = -J\sum_{\langle ij \rangle} \cos(\theta_i-\theta_j) - \gamma_a\sum_i \cos^2\theta_i - h_x^{\text{ext}}(t)\sum_i \cos\theta_i,
\end{equation}
where $\theta_i \in [0,2\pi)$. For $\gamma_a > 0$ , the system prefers to be aligned along the $(\pm)~x$-axis while for $\gamma_a<0$ the system prefers alignment along the $(\pm)~y$-axis. For the time dependent external magnetic field, we have used a square wave drive protocol of amplitude $h_0$ and frequency $\Omega$. 
\subsection{Monte Carlo method \label{sec:montecarlo}}
\subsubsection{\label{subsec:algo}Algorithm}

The interacting spins in our model are driven by an external magnetic field $h_x^{\text{ext}}(t)$, while also being in contact with a heat reservoir at inverse temperature $\beta=1/k_B T$, where $k_B$ is the Boltzmann constant and $T$ is the temperature. In the rest of the article, we fix $\beta=4$ such that the temperature is sufficiently low to ensure ordering in equilibrium, at least at short length-scales, for a non-zero $\gamma_a$. It has been shown in previous studies (see \cite{LoPhysRevA.42.7471,RikvoldPhysRevE.49.5080,AcharyyaPhysRevE.56.2407,AcharyyaPhysRevE.56.1234,KornissPhysRevE.63.016120,FujisakaPhysRevE.63.036109,RaoPhysRevB.42.856,SidesPhysRevE.57.6512}) that MC methods, such as the single spin flip Glauber algorithm \cite{Glauber}, captures the essential microscopic dynamics of such systems. The elementary move of the MC algorithm consists of updating the spin at site $i$ from $\vec{S}_i$ to some randomly proposed orientation $\vec{S}'_i$. This proposed move will be accepted with the probability $W_{\text{acc}}=1/\Big(1+e^{\beta\Delta E_i}\Big)$, which satisfies the instantaneous detailed balance condition. It is important to note that $\Delta E_i$ is the energy difference of the proposed and the current spin configuration computed using the instantaneous external magnetic field $h_{x}^{\text{ext}}(t)$, which keeps changing with MC time. For a $L\times L$ lattice, attempting $L^2$ such elementary moves constitutes one MC step per spin (MCSS), which serves as the \textit{unit} of time for the microscopic dynamics of the spins. For a square wave drive of frequency $\Omega_{\text{MC}}$ the external magnetic field $h_{x}^{\text{ext}}(t)$ switches direction after time $T_{\text{MC}}/2$, where $T_{\text{MC}}=2\pi/\Omega_{\text{MC}}$. The label $\text{MC}$, in $\Omega_{\text{MC}}$, emphasizes that time is measured in units of MCSS. We define a full MC cycle to be $T_{\text{MC}}$ MCSS.

Following the usual route in MC simulation of equilibrium statistical mechanics, we first initialize the spin configuration and then update it according to the MC moves described above. In all our MC calculations, we have used a random initialization of the spins. The late time steady state values of $|\langle Q_x \rangle|$ and $|\langle Q_y \rangle|$ for different initial conditions should agree with each other within error bars. We have verified this to be true for three different initial conditions, namely, (i) random, (ii) $X$ ordered ($\vec{S}_i=\left(1,0\right),\forall i$), and (iii) $Y$ ordered ($\vec{S}_i=\left(0,1\right),\forall i$), for several choices of $h_0,\gamma_a$, and $\Omega_{\text{MC}}$. The system goes through a transient dynamics, and eventually after a sufficiently large number of cycles, reaches a quasi-periodic steady state. The trajectory at late times, is not strictly periodic, due to the presence of thermal fluctuations supplied by the heat reservoir. However,the magnitudes of the order parameters $|Q_x|$, $|Q_y|$ attain steady state values. We now discuss the key observables that we monitor in order to analyze the DPTs in this system. 

\subsubsection{\label{subsec:observ}Observables}

The microscopic simulation using the MC method described above, allows us to directly probe the magnetization of the system along the $x$ or the $y$ axes (in spin space) for an individual spin configuration. When averaged over one complete cycle, this gives us the dynamical order parameters $Q_x,Q_y$ which in turn characterizes the dynamical phase of the system (see Table \ref{tab:table1} for definition of the four possible dynamical phases). For a spin configuration having angles $\{ \theta_i(t_{\text{MC}}) ,i \in [1,L^2]\}$ at time $t_{\text{MC}}$ (in units of MCSS) the $x,y$ magnetizations are given by \begin{subequations}
\label{eq:mc-magnetizations}
 \begin{align}
      m_x(t_{\text{MC}}) = \frac{1}{N}\sum_{i=1}^{N} \cos[\theta_i(t_{\text{MC}})],\\
      m_y(t_{\text{MC}}) = \frac{1}{N}\sum_{i=1}^{N} \sin[\theta_i(t_{\text{MC}})],
\end{align}
\end{subequations} where $N=L^2$ is the total number of sites of the lattice. The various moments of the dynamical order parameters are the time averages of the corresponding moments of $x$ and $y$ magnetization components over one complete cycle having a temporal width $T_{\text{MC}}$. The expression for the average of the $r^{th}$ moment of $Q_{\alpha}$ is given by
\begin{equation}
    \label{eq:mc-orderparameters}
      Q^{(r)}_{\alpha} = \frac{1}{T_{\text{MC}}}\sum_{t_{\text{MC}}=1}^{T_{\text{MC}}}m^r_{\alpha}(t_{\text{MC}}), 
\end{equation}
where $\alpha=x,y$ and $r$ is a positive integer. For $r=1$, these are the dynamical order parameters $Q_x$ and $Q_y$, while the higher moments are used to define the dynamic analogs of magnetic susceptibilities and the Binder cumulants.

In order to verify the universality class of the DPTs between the different dynamical phases listed in Table~\ref{tab:table1}, we have carried out the FSS analysis of the susceptibility and the fourth order Binder cumulant across these transitions. The susceptibility and the Binder cumulants corresponding to the dynamical order parameters $Q_x$ and $Q_y$, for a system of linear size $L$, are given by
\begin{subequations}
\begin{align}
\label{eq:susc-defn}
\chi_{\alpha}\left(L\right) &= L^2\beta( \langle Q_{\alpha}^2 \rangle_{L} - \langle |Q_{\alpha} |\rangle_{L} ^2),\\
\label{eq:U4-defn}
U_\alpha^{\left(4\right)} \left(L\right)&= 1-\frac{\langle Q_{\alpha}^4 \rangle_L}{3\langle Q_{\alpha}^2 \rangle^2_L}.
\end{align}
\end{subequations} 

In the above expression $\langle~...~\rangle$ denotes average over several MC cycles after the initial transient has passed. Symmetry arguments suggest that, if across any transition, the nature of \textit{exactly one} order parameter changes, then that transition is of the Ising universality class if it is continuous. It should be noted however that the resulting Ising transition is triggered by the magnetic field and not temperature (which is kept  to $T=1/4$ during the simulations). Hence one must use the appropriately modified scaling relations for susceptibility and the Binder cumulants given by \begin{subequations}
\begin{align}
   \label{eq:susc-scaling}
   \chi_{\alpha}(g,L) L^{-\gamma/\nu}& = f_1(gL^{1/\nu}), \\
   \label{eq:U4-scaling}
   U^{\left(4\right)}_{\alpha}\left(g,L\right) & = f_2(gL^{1/\nu}),
\end{align}
\end{subequations} where $\alpha=x,y$ and $f_1,f_2$ are the same universal functions describing the order-disorder 2D Ising thermal phase transition. The critical exponents $\gamma,\nu$ have the well known values $\gamma=7/4,\nu=1$ with $h^\star $ being the critical point and $g=h/h^{\star}-1$ which can be computed by performing a FSS collapse \cite{10.1063/1.3518900}. All the MC results were obtained through simulations of a range of lattice sizes, $L=64,128,150,180,200,256,512,1024$. For system sizes $L\le 256$, we simulated a total of 3$\times 10^6$ MC cycles, of which the initial 1/3 cycles were used for equilibration and the remaining 2/3 were used for measurements and analysis. Similarly for system sizes $L=512$ and $1024$, we have simulated a total of 3$\times 10^7$ MC cycles , of which the initial 1/3 cycles were used for equilibration and the remaining 2/3 were used for measurements and analysis. Earlier studies of the 2d NNKI using stochastic Glauber dynamics found evidence~\cite{KornissPhysRevE.63.016120, SidesPhysRevLett.81.834} of the DPT to be identical to the 2d Ising universality class in equilibrium, and the same is expected for the 2d an-$XY$ model for DPTs between Ising-SBO and Ising-SRO if the transition is continuous, which we have verified from our MC simulations. However, the fate of the DPTs between two dynamical phases where at least one of them is non-Ising in nature might show new features and we have addressed this carefully using FSS in the next section.    

The simulations were performed on GPGPUs (general purpose graphics processing units) using the heterogeneous (CPU+GPU) computing paradigm CUDA \cite{cuda,RuetschCUDA}. We have implemented the Glauber MC algorithm \cite{Glauber} on the GPU using the heterogeneous (CPU-GPU) programming model CUDA  \cite{RuetschCUDA}. For the MC simulations we utilized a CUDA implementation of the RANLUX pseudo-random number generator algorithm which directly generates random numbers on the GPU, bypassing the need for host (CPU) to device (GPU) data transfer. Moreover, exploiting the first nearest neighbour coupling of the model, the MC update procedure can be readily parallelized using the standard checkerboard decomposition of a two dimensional square lattice. 
For example, on an NVIDIA Tesla-V100 GPU, using a one dimensional grid configuration of $512$ blocks and $12$8 threads per block, we can associate all sites of a $256\times256$ lattice with one dedicated GPU thread. Since a MC update on each site is a relatively simple process, it is extremely well suited for the GPU environment. We can offload the update of an entire sub-lattice ($32768$ sites for the above example) to the GPU. For this system size performing $10^4$ updates takes approximately $\sim 30$ seconds. We note that it may be possible to generalize advanced GPU optimization schemes, already available for Ising spins~\cite{ROMERO2020107473}, to the case of continuous spins for even more efficient implementation of the MC simulations. Next, we compute the observables after a desired number of MC updates have been performed on the entire lattice. We record statistics of the observables after every MC sweep and this process is continued as long as necessary for reaching a desired level of accuracy. 
  
\subsection{Mean-field theory \label{sec:mft}} 
We now present the formulation of a MF approach, which reproduces several aspects of the exact MC phase diagram. In this approach,  starting from the microscopic Hamiltonian (Eq.~\eqref{eq:hamXY}), we construct a set of \textit{phenomenological} EOMs which enable us to study the dynamical features of Eq~\eqref{eq:hamXY} at the level of MF approximation. The derivation of these MF EOMs proceeds as follows. First, as per the standard MF approximation, we assume the fluctuations to be small which enable us to write $\vec{S}_i \approxeq \vec{m}+\delta\vec{m}$ , neglecting terms of order $\mathcal{O}(\delta\vec{m}^2)$. We can now express the nearest neighbor ferromagnetic coupling term in Eq.~\eqref{eq:hamXY} as 
\begin{equation}
        -J\sum_{\langle ij \rangle} \vec{S}_i \cdot \vec{S}_j \approxeq \frac{JNq\vec{m}^2}{2} - Jq\vec{m}\cdot \Big( \sum_i \vec{S}_i \Big),
        \label{eqn:MF_approx}
\end{equation}
where $q$ is the coordination number of the lattice. For the case of a $2d$ square lattice, in which we are interested, $q=4$. Using Eq.~\eqref{eqn:MF_approx}, the resulting total MF Hamiltonian becomes a sum of single site terms leading to the MF partition function given by
\begin{equation}
\begin{split}
\mathcal{Z}_{\text{MF}} & = \exp\Bigg(\frac{\beta JNq \vec{m}^2}{2}\Bigg)\int_{0}^{2\pi} \prod_r d\theta_r\; \exp\Bigg( \beta \vec{h}_{\text{eff}}(t) \cdot \sum_i \vec{S}_i \Bigg)  \\ & \;  \qquad\qquad\qquad\qquad\qquad \exp\Bigg( \beta \gamma_a \sum_i( \vec{S}_i\cdot\hat{x})^2 \Bigg),\\
& \propto  \Bigg[ \int_{0}^{2\pi} d\theta  \exp\Big( \beta\gamma_a \cos^2\theta \Big) \exp\Big( \beta h_{\text{eff}}\cos(\theta-\phi) \Big) \Bigg]^N,
\end{split}
\label{eq:MF-partition}
\end{equation}
where for convenience of notation we have defined the effective magnetic field $\vec{h}_{\text{eff}}= Jq\vec{m} + \vec{h}_{\text{ext}}(t) $. Its component in the $x$ ($y$) direction is defined as $h_{\text{eff}}^x=h_{\text{eff}}\cos\phi ~(h_{\text{eff}}^y = h_{\text{eff}}\sin\phi)$, respectively and the orientation $\phi$ relative to the $x$ axis is given by $\tan\phi = h_{\text{eff}}^y/ h_{\text{eff}}^x$. It is worthwhile to note that unlike $J\sum_{\langle ij \rangle} (\vec{S}_i \cdot \vec{S}_j$), the other two terms in Eq.~\eqref{eq:hamXY} do not couple to different spins and thus do not require any application of the MF approximation. By expressing the integrand in the RHS of Eq.~\eqref{eq:MF-partition} in terms of modified Bessel functions, one can explicitly perform the configuration space integration. This, however, comes at the cost of dealing with infinite sums of modified Bessel functions of all orders. After performing the said integral (see Sec.~\ref{appendix-derivation-ZMF}), modulo some proportionality factors, we find the following expression of the MF partition function 
\begin{equation}
\label{eq:MF-partition-simplified}
\mathcal{Z}_{\text{MF}} \propto \exp\Big(\frac{\beta JNq\vec{m}^2}{2}\Big) \Bigg[ \sum_{\mu=-\infty}^{\infty} \mathcal{I}_{2\mu}(\beta h_{\text{eff}})\: \mathcal{I}_{\mu}(\beta\gamma_a/2)e^{2i\mu\phi}\: \Bigg]^N .
\end{equation}
The corresponding free energy (per site) is then given by
\begin{equation}
\label{eq:free-energy}
\begin{split}
f_{\text{MF}} & = -\frac{1}{N}k_BT\ln\mathcal{Z}_{\text{MF}} \\
          & = \frac{Jq\vec{m}^2}{2}-\frac{1}{\beta}\ln\left[ \sum_{\mu=-\infty}^{\infty} \mathcal{I}_{\mu}(\beta\gamma_a/2)\mathcal{I}_{2\mu}(\beta h_{\text{eff}}) \cos(2\mu\phi) \right].
        \end{split}
\end{equation}

We are now in a position to introduce the MF EOMs. From Eq.~\eqref{eq:free-energy} one can define a set of \textit{phenomenological} coupled EOMs. Such equations were used previously within the context of the Ising model \cite{TomePhysRevA.41.4251} and are based upon the simple intuition that in an out-of-equilibrium scenario, the system continually tries to minimize its free energy. This condition can be satisfied if we demand that the system flows along the local gradient of the free energy. Adapting this idea to our context, we arrive at the following set of coupled differential equations \begin{subequations}
\label{eq:XYdmdt}
\begin{align}
\Gamma_x \frac{dm_x}{dt}=-\frac{\partial f_{\text{MF}}}{\partial m_x}=-Jqm_x + G_x(m_x,m_y), \\
        \Gamma_y \frac{dm_y}{dt}=-\frac{\partial f_{\text{MF}}}{\partial m_y}=-Jqm_y + G_y(m_x,m_y).   
\end{align}
\end{subequations}
In the above $\Gamma_x$ and $\Gamma_y$ are the phenomenological friction or damping coefficients along the $x$ and $y$ directions, respectively. In general, it is possible that $\Gamma_x \ne \Gamma_y$. However, for simplicity we have set $\Gamma_x=\Gamma_y=1$. $G_x(m_x,m_y)$ and $G_y(m_x,m_y)$ contain terms arising from the derivatives of the modified Bessel functions. The explicit forms of $G_x(m_x,m_y)$ and $G_y(m_x,m_y)$ can be found in Eq.~\eqref{eq:appendix-GxGy}. 

By solving the coupled differential equations Eq.~\eqref{eq:XYdmdt} numerically, one can find the MF trajectory in the $m_x-m_y$ plane. This in turn will allow us to compute the dynamical order parameters.  We used the standard library ODEINT (\cite{2020SciPy-NMeth}) to numerically integrate these EOMs. We use the default settings for the LSODA solver from ODEPACK where relative tolerance is $10^{-3}$, absolute tolerance is $10^{-6}$. For each half cycle of the drive, the external field $h_x^{\text{ext}}$ is  either to $+h_0$ (positive half) or to $-h_0$(negative half). The Eqs.~\eqref{eq:XYdmdt} are evolved with the respective (constant) value of $h_x^{\text{ext}}$ for every half-cycle of the drive. Much like the MC trajectories, these MF trajectories also have an initial transient. The transient depends on the choice of the initial condition. After a sufficient number of cycles, the transient passes and the system chooses a stable orbit with the same periodicity as the driven external magnetic field. 

Once the trajectory is found, one can compute the dynamical order parameters $Q_x$ and $Q_y$ given by $Q_\alpha = \frac{\Omega_{\text{MF}}}{2\pi} \oint m_\alpha(t)\; dt$ (where recall $\alpha =x, y$). This enables us to label the dynamical phase following Table~\ref{tab:table1}. The above integrals are taken over one complete drive cycle (after stabilization of the orbits), and then averaged over several cycles. When comparing the results of the MC and the MF methods, it should be kept in mind that their time scales are related to each other by some unknown scaling factor. For this reason a direct quantitative comparison of the MF and the MC phase diagrams is difficult, if not impossible, and is not attempted here. To emphasize this difference of timescales, different labels have been used while specifying the frequencies ($\Omega_{\text{MC}}$ for MC and $\Omega_{\text{MF}}$ for MF) throughout this article. All the other parameters such as anisotropy $\gamma_a$, magnetic field $h_0$ as well as the temperature $T$ are exactly equivalent in the two methods owing to the derivation of the MF free energy per site (Eq.~\eqref{eq:free-energy}) directly from the microscopic Hamiltonian (Eq.~\eqref{eq:hamXY}). The specific values of frequencies in the two methods where chosen to have the transitions occur within the same order of magnitude of the magnetic field.  In the next section, we utilize MF and MC methods described in this section to compute and analyze the dynamical phases and the transitions between them.

\begin{figure*}[!htbp]
\centering
\includegraphics[width=0.9\textwidth]{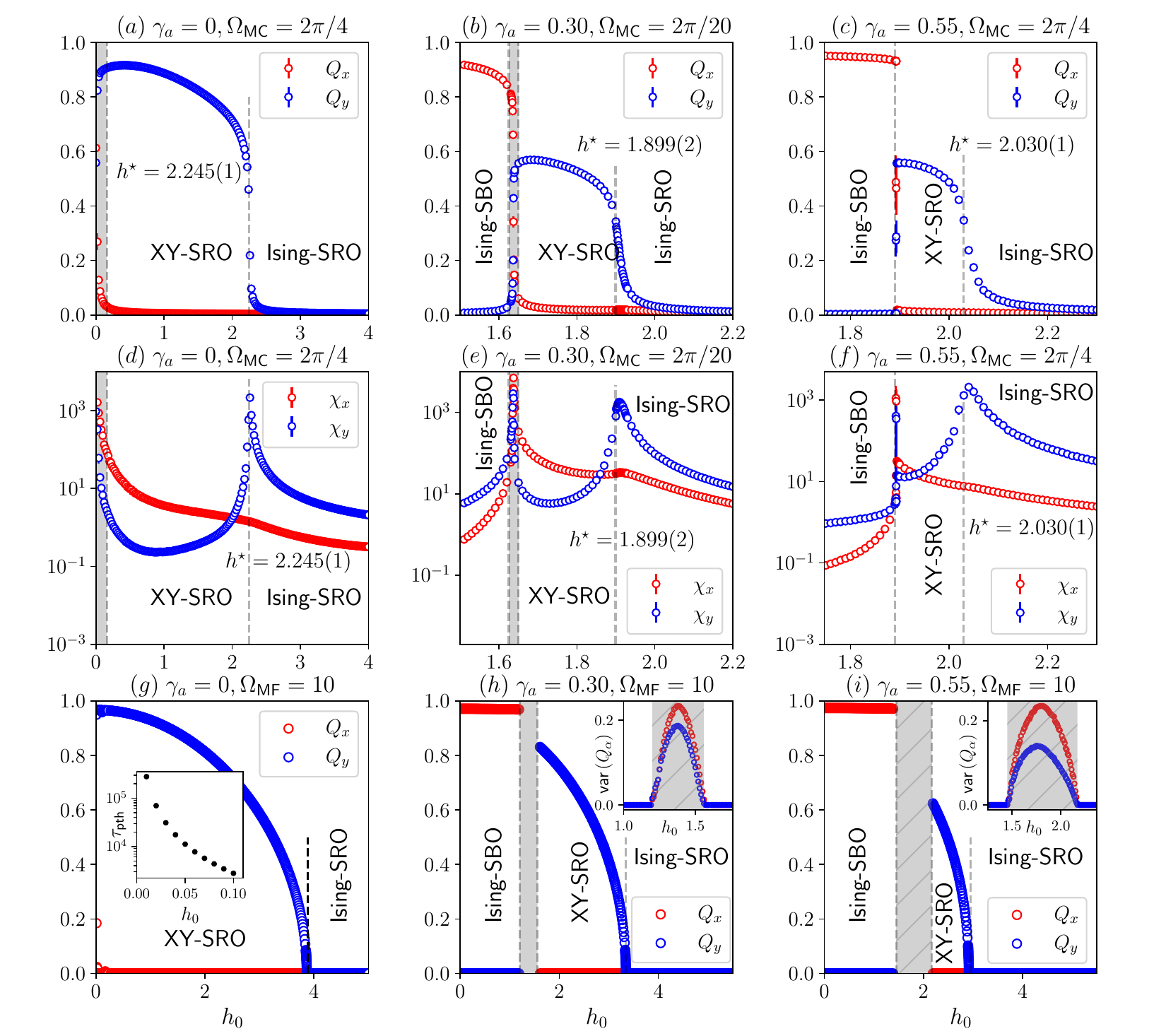}
\caption{Phase diagrams obtained from Monte Carlo and mean-field calculations for different parameter choices. Upper panels (a)-(c) show MC data for the order parameters computed using Eqs.~\eqref{eq:mc-magnetizations} and \eqref{eq:mc-orderparameters} for lattice size $L=200$ while the middle panels (d)-(f) show MC data for the magnetic susceptibilities computed using Eq.~\eqref{eq:susc-defn} using the same parameter sets as in (a)-(c). The lower panels (g)-(i) are mean-field calculations where the order parameters are obtained from the numerical solutions of Eq~\eqref{eq:XYdmdt}. The unit of time is different in the two calculations, see Secs.~\ref{sec:montecarlo} and ~\ref{sec:mft} for discussion. Critical values of the magnetic field $h^{*}$ are obtained using a FSS analysis as detailed in Sec.~\ref{subsec:$XY$dpt} and denoted by vertical dashed lines in panels (a)-(f). The critical fields were computed using a finite size scaling analysis, see Figs.~\ref{fig:fig3}(e) and \ref{fig:fig3}(f), using the formulae described in Eqs.~\eqref{eq:susc-scaling},\eqref{eq:U4-scaling}. All Monte Carlo simulations were performed at an inverse temperature of $\beta=4$ for reasons mentioned in Sec.~\ref{subsec:algo}. The definition of the dynamical phases are defined in Table~\ref{tab:table1}. The inset in panel (g) shows the number of cycles required for reaching the $XY$-SRO steady state in the presence of small magnetic field strength. The insets in panels (h) and (i) highlight the sensitivity towards initial condition dependence of the late time phase in the respective magnetic field regime by showing the behaviour of the variance of the order parameters $\text{var}\left(Q_x\right)$ (red points) and $\text{var}\left(Q_y\right)$ (blue points) for a uniform grid of $100$ initial conditions with $m_{x,y} \in [-1,1]$ as a function of $h_0$. See Sec.~\ref{sec:results} for discussion.}
\label{fig:fig1}
\end{figure*}

\section{\label{sec:results}Results} 

We now present our findings on the dynamical phases, their stability as analyzed from both MC and MF calculations, and the FSS analysis of the phases and the transitions from the MC data for different choices of parameters. 

\subsection{\label{subsec:$XY$dpt}Dynamic Phases and Phase Transitions}
In Figs.~\ref{fig:fig1}(a)-(f), we discuss the findings of our MC simulations for a fixed lattice size $L=200$. In Fig.~\ref{fig:fig1}(a) we show the phase diagram for zero anisotropy ($\gamma_a=0$) at $\Omega_{\text{MC}}=2\pi/4$ for an external magnetic field amplitude of $h_0$, where the time period of the system is $T_{\text{MC}}=4$. The phase diagram shows that there is a thin sliver of $XY$-SBO , where both $|Q_x|$ and $|Q_y|$ are non-zero, which exists very close to the zero field condition. Upon increasing the field, even slightly (indicated by the vertical dashed line in Fig.~\ref{fig:fig1}(a)), the $XY$-SBO phase vanishes rapidly and changes into a $XY$-SRO phase. Increasing the magnetic field further, we observe a phase transition from $XY$-SRO to Ising-SRO. The second transition occurs at a critical field of $h^{*}=2.245$. The values of the critical couplings $h^{*}$ reported in this article were obtained from FSS using the method outlined in Sec.~\ref{sec:montecarlo} and explained in more detail later in this subsection. The phase diagram also shows that in the absence of anisotropy, the $Q_y$ order parameter dominates the system. Our MC calculations suggest that in the thermodynamic limit ($L\rightarrow\infty$) and for $\gamma_a \leq 0$ the only existing phases are Ising-SRO and $XY$-SRO. This finding is consistent with Ref.~\onlinecite{YasuiPhysRevE.66.036123}. 

\begin{figure*}[ht]
    \centering
    \includegraphics[width=0.9\textwidth]{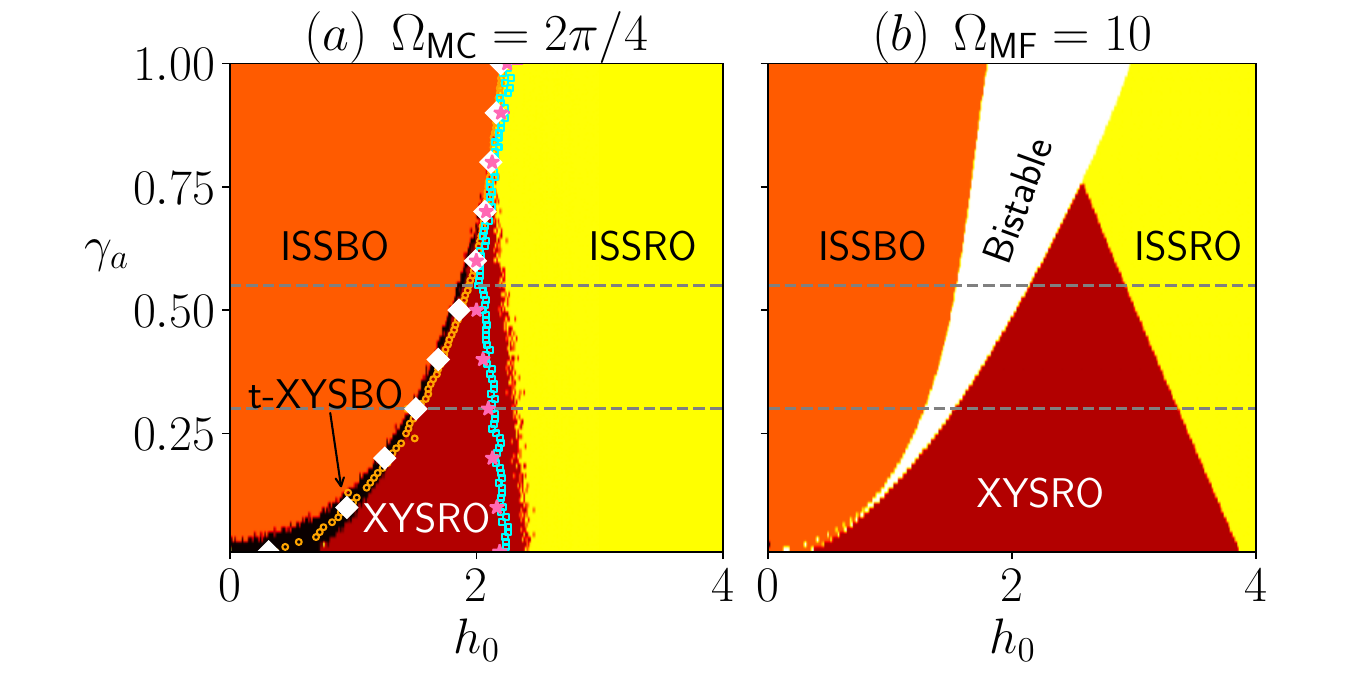}
    \caption{Phase diagram in the $\gamma_a-h_0$ plane computed using $(a)$ Monte Carlo simulations for $L=64$ and $(b)$ mean-field method. To generate the phase diagram in $(a)$ we have used an operational definition of taking $\langle |Q_{x}|\rangle$ or $\langle |Q_{y}|\rangle$ less than 0.05 to be effectively zero given the finite size of the lattice. The reasonableness of this cutoff value has been checked by simulating $L=256$ for specific values of $\gamma_a$. The t-$XY$-SBO region refers to the ``temporary" $XY$-SBO region that is a finite-size effect and vanishes as 
    $L \rightarrow \infty$ in (a). A more accurate method to locate the phase boundaries is provided by tracking the peak positions of $\chi_x$ and $\chi_y$ as a function of $h_0$ for different values of $\gamma_a$. The open orange circles (cyan squares) are the peak locations of susceptibilities $\chi_x$ ($\chi_y$) for $L=64$. The filled white squares (pink stars) are the peak locations of $\chi_x$ ($\chi_y$) for $L=256$. The bistable region (white) in (b) refers to the case where generic initial conditions can be grouped in two subsets where each subset flows to a unique dynamical phase characterized by the magnitude of its order parameters $(|Q_x|,|Q_y|)$. The subsets can flow to (i) either Ising-SBO or $XY$-SRO or to (ii) either Ising-SBO or Ising-SRO depending on the specific parameter values inside the bistable region. The gray dashed lines in both panels correspond to 1d cuts shown in Fig.~\ref{fig:fig1}(a)-\ref{fig:fig1}(i).}
    \label{fig:fig2}
\end{figure*}

We investigate the system for finite positive anisotropy where the choice of $\gamma_a = 0.30$ and $0.55$ were based on the $\gamma_a - h_0$ phase diagram plot shown in Fig.~\ref{fig:fig2}. In Fig.~\ref{fig:fig1}(b) we show the phase diagram for $\gamma_a=0.30$ and $\Omega_{\text{MC}}=2\pi/20$, where the time period is $T_{\text{MC}}=20$. For these parameter combinations we find all four dynamical phases and three phase transitions between them. The shaded region represents the $XY$-SBO phase. At each transition only one order parameter dominates, either $Q_x$ or $Q_y$. For example, in the transition between the $XY$-SBO phase and the $XY$-SRO phase, $Q_y$ becomes non-zero, but $Q_x$ almost vanishes. The small non-zero value of $|Q_x|$ observed in the $XY$-SRO region of Fig.~\ref{fig:fig1}(b) is a finite-size effect, and $|Q_x|\rightarrow 0$ in the limit of large lattice sizes. This implies that all the DPTs must be of the Ising universality class as long as they are continuous from symmetry considerations. Furthermore, we note that in the presence of finite anisotropy ($\gamma_a\ne0$), the $XY$-SBO zone shifts from the vicinity of the zero field to a finite magnetic field value of around $h_0 \approx 1.62$. We also find that the transition boundary between the $XY$-SRO and the Ising-SRO phase now occurs at a downshifted value of $h^{*}=1.899$. Figs.~\ref{fig:fig1}(a) and \ref{fig:fig1}(b) suggest that the $XY$-SBO phase is typically present in a very narrow region of magnetic field, at least for the parameter regimes explored here. This raises the obvious question whether the $XY$-SBO phase is stable in the thermodynamic limit ($L\rightarrow\infty$) or not. To test finite size effects on the occurence of $XY$-SBO state, we considered lattice sizes much larger than $L=200$. We conclude that the presence of $XY$-SBO in the MC simulation is a finite size feature which vanishes in the thermodynamic limit. 

In Fig.~\ref{fig:fig1}(c) we show the phase diagram for $\gamma_a=0.55$ and $\Omega_{\text{MC}}=2\pi/4$, where the time period is $T_{\text{MC}}=4$. We find that at this value of anisotropy, the $XY$-SBO phase is completely absent. In the transition between Ising-SBO ($Q_x\ne0,Q_y=0$) and $XY$-SRO ($Q_x=0,Q_y\ne0$), the value of both dynamical order parameters, $Q_x$ and $Q_y$ interchange. The values switch from a non-zero to a zero value and vice versa. Such a transition must be first order in nature. The MC results also suggest that further increasing the anisotropy can lead to a direct transition between the Ising-SBO and Ising-SRO phase. This behavior can be observed both from the MF and MC calculations, see Fig.~\ref{fig:fig2}. 

We show the behavior of the corresponding magnetic susceptibilities $\chi_x$ and $\chi_y$ (see Eq.~\eqref{eq:susc-defn}) for the parameter values used in Fig.~\ref{fig:fig1}(a)-(c) in Fig.~\ref{fig:fig1}(d)-(f) to clarify the magnetic response of the different dynamical phases. The magnetic susceptibilities display prominent peaks (the y-axis is displayed in logarithmic scale for Fig.~\ref{fig:fig1}(d)-(f)) in the neighborhood of the critical magnetic fields both for the transitions between Ising-SBO to $XY$-SRO as well as between $XY$-SRO to Ising-SRO as $h_0$ is increased. While only $\chi_y$ shows a peak for the transition between $XY$-SRO to Ising-SRO, both $\chi_x$ and $\chi_y$ display peaks as the system crosses from Ising-SBO to $XY$-SRO as $h_0$ is varied. The values of the critical coupling $h^*$ in the thermodynamic limit for the $XY$-SRO to Ising-SRO transition obtained using FSS is seen to be close to the location of second peak of $\chi_y$ at a higher $h_0$ already for $L=200$.

Our observation of three dynamical phases: Ising-SBO, $XY$-SRO and Ising-SRO from MC calculations for $\gamma_a=0.55$ does not contradict the behavior predicted by Ref.~\onlinecite{YasuiPhysRevE.66.036123}. Based on a time dependent Landau Ginzburg analysis they concluded that for values of $\gamma_a>1/2$ only Ising-SRO and Ising-SBO phases can exist. We note that our MC plots were computed for a higher frequency (lower time period) than Yasui \emph{et al}. Thus, a direct quantitative comparison should not be made. Based on the value of the time-period in the system the critical $\gamma_a$ value can alter the boundary of the direct Ising-SRO to $XY$-SBO transition. From this perspective $\gamma_a>1/2$ is a soft constraint. Even within our calculation we can conclude that for higher frequencies there will be a critical value of $\gamma_a$ above which a direct transition can exist.

\begin{figure}[!hbtp]
\centering
\rotatebox{0}{\includegraphics*[width= 0.90 \linewidth]{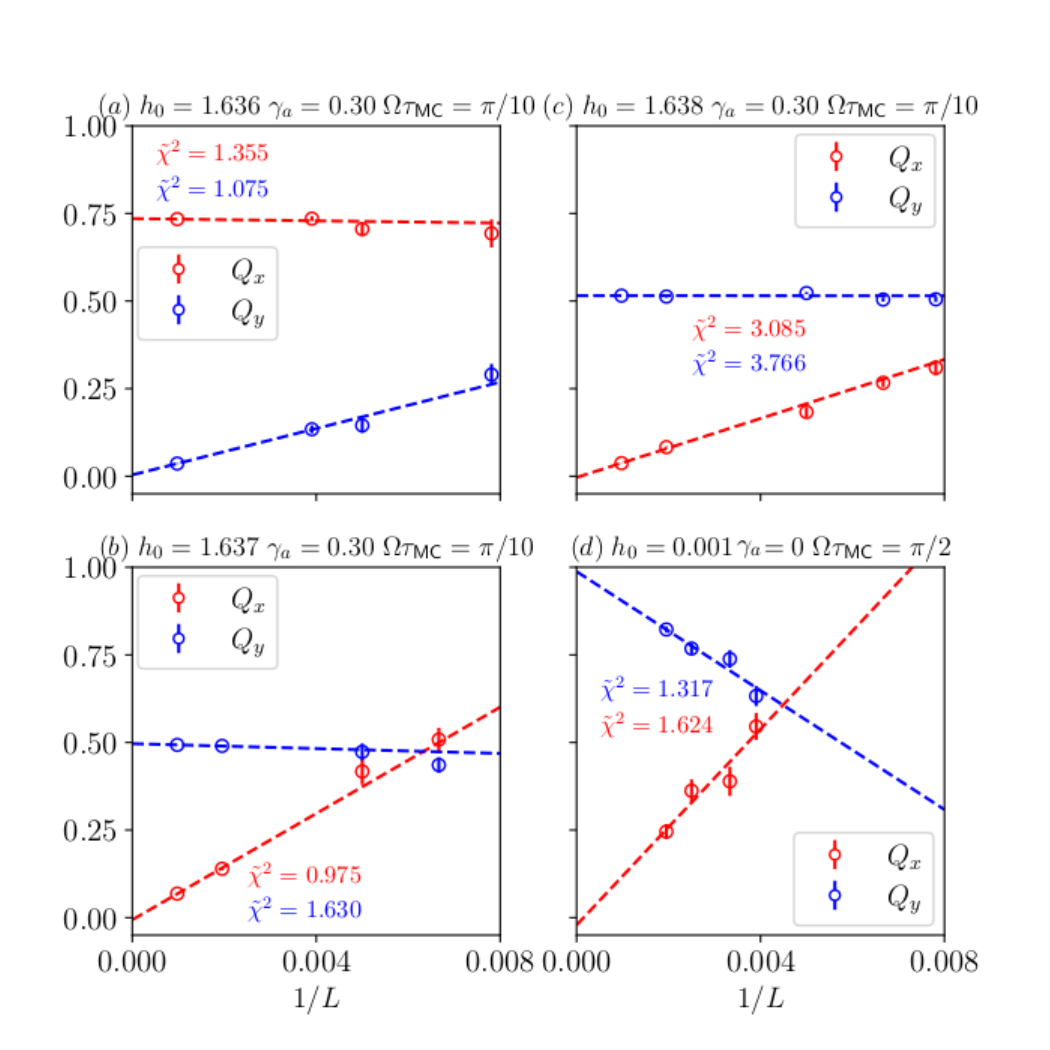}}
\rotatebox{0}{\includegraphics*[width= 0.90 \linewidth]{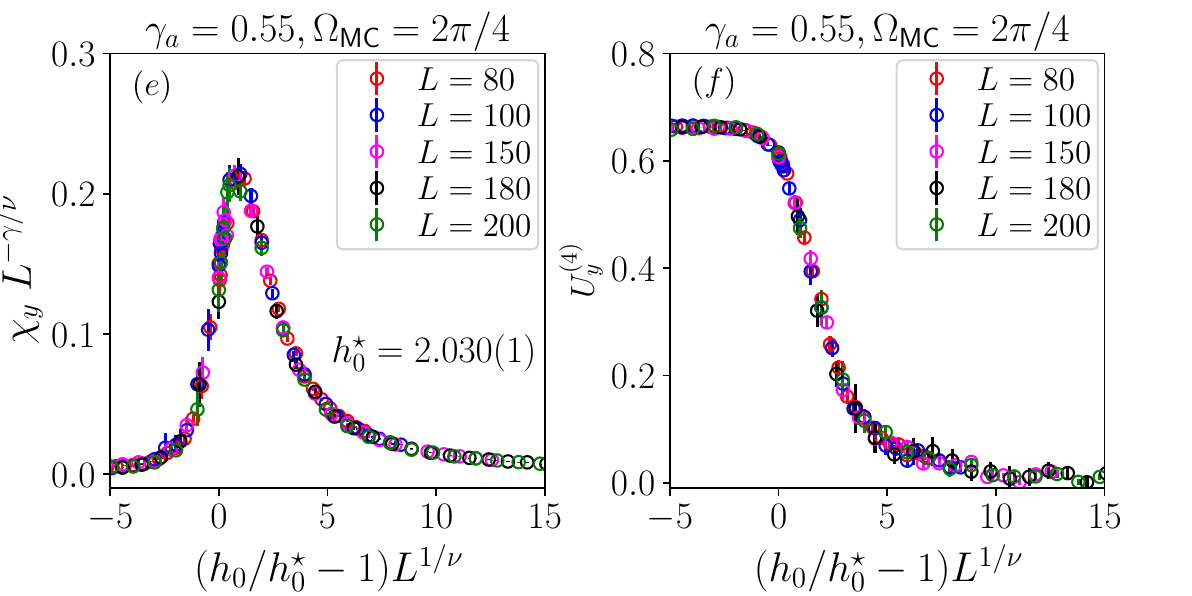}}
\caption{Panels (a)-(d) show the finite size extrapolation of the absolute values of order parameters $|Q_x|$(red) and $|Q_y|$(blue) inside the narrow $XY$-SBO regions in Figs.~\ref{fig:fig1}(a)-\ref{fig:fig1}(c) to larger lattices. In panels $(a)-(d)$ we show the chi-squared per degree of freedom $\tilde\chi^2$ obtained during the fitting of the dashed straight lines. Our results indicate that as we approach the thermodynamic limit, the $XY$-SBO phase becomes destabilized. In panels (e) and (f) we verify the universality class of the $XY$-SRO to Ising-SRO transition. We have calculated the variation of $\displaystyle y_L = \chi_y L^{-\gamma/\nu}$ against $\displaystyle x_L = gL^{1/\nu}$ in $(e)$ and $\displaystyle y_L =U_{y}^{\left(4\right)}$ against $x_L$ in $(f)$ where $g=h/h^\star-1$.  The figures show that the transition at $\gamma_a=0.55, \Omega_{\text{MC}}=2\pi/4$ falls in the $2d$ Ising universality class. Similar scaling collapses have been performed for $\:\:\gamma_a=0,\Omega_{\text{MC}}=2\pi/4$ and $\:\:\gamma_a=0.30,\Omega_{\text{MC}}=2\pi/20$. Based on these calculations we can confirm that the $XY$-SRO to the Ising-SRO transition is always Ising like. This allowed us to find the critical values of the magnetic fields as noted in Figs.~\ref{fig:fig1}(a) and \ref{fig:fig1}(c), respectively. The scaling collapse is performed by taking the $2d$ Ising critical exponents $\nu=1$ and $\gamma=7/4$. The scaling forms are described in Eqs.~\eqref{eq:susc-scaling} and \eqref{eq:U4-scaling}.  }
	\label{fig:fig3}
\end{figure}

In Figs.~\ref{fig:fig1}(g)-\ref{fig:fig1}(i) we show the corresponding phase diagram obtained from MF calculations. For $\gamma_a=0$ and small $h_0$ (Fig.~\ref{fig:fig1}(g)), the MF EOMs predict a long prethermal regime that mimics $XY$-SBO in that $|Q_x|$ and $|Q_y|$ are both non-zero before $|Q_x|$ eventually decays to zero as $\exp(-t_{\mathrm{MF}}/\tau_{\mathrm{pth}})$, where $\tau_{\mathrm{pth}}$ is the pre-thermal lifetime. The behavior of $\tau_{\mathrm{pth}}$ as a function of $h_0$ is shown in the inset of Fig.~\ref{fig:fig1}(g). The grey regions marked in the main panels of Figs.~\ref{fig:fig1}(h)-\ref{fig:fig1}(i) show the presence of a ``bistable region'' from the MF EOMs where while a class of initial conditions eventually lead to a periodic steady state with Ising-SBO and identical value of $|Q_x| \neq 0$ (independent of the initial condition), other initial conditions lead to $XY$-SRO with identical value of $|Q_y| \neq 0$. The insets of Fig.~\ref{fig:fig1}(h) and Fig.~\ref{fig:fig1}(i) show the variance of both $|Q_x|$ and $|Q_y|$ calculated using several different initial conditions (100 initial conditions on a uniform grid formed by $m_x\in [-1,1]$ and $m_y\in[-1,1]$) at each parameter value. For parameter values displayed in Figs.~\ref{fig:fig1}(g)-\ref{fig:fig1}(i) that are outside this bistable region, generic initial conditions lead to the same value of dynamical order parameters $|Q_x|$ and $|Q_y|$ at late times, independent of the initial condition.

In Fig.~\ref{fig:fig2}, we show the phase diagram for the dynamical phases in the $\gamma_a-h_0$ plane for $\Omega_{\mathrm{MC}}=2\pi/4$ from MC simulations at a fixed system size of $L=64$ (Fig.~\ref{fig:fig2}(a)) and for a fixed drive frequency of $\Omega_{\mathrm{MF}}=10$ from MF calculations (Fig.~\ref{fig:fig2}(b)). While the MF phase diagram explicitly shows the three stable dynamical phases, namely, $XY$-SRO, Ising-SBO and Ising-SRO, apart from a region of bistability, the MC phase diagram is based on an operational definition of taking $|Q_x|$ or $|Q_y|$ less than $0.05$ to be effectively zero given the finite size of the lattice. A better, but more computationally intensive method, to locate the phase boundaries is to use the peak positions of 
$\chi_x$ and $\chi_y$ by varying $h_0$ for several values of $\gamma_a$ for a given $L$ to define pseudo-critical couplings that converge to the critical couplings as $L \rightarrow \infty$. The open symbols in Fig.~\ref{fig:fig2}(a) are from such an analysis of $\chi_x$ and $\chi_y$ for a system size of $L=64$ for many values of $\gamma_a$ while the filled symbols are obtained using the magnetic susceptibility data for a bigger lattice of $L=256$ for a smaller number of $\gamma_a$ values. Comparing peak positions shows that the shift in the phase boundaries obtained using this method is small when one increases the system size from $L=64$ to $L=256$. Based on the plots, we can conclude that the MF method adequately reproduces the MC simulation results.

\begin{figure*}[!ht]

  \rotatebox{0}{\includegraphics[width=0.93\linewidth]{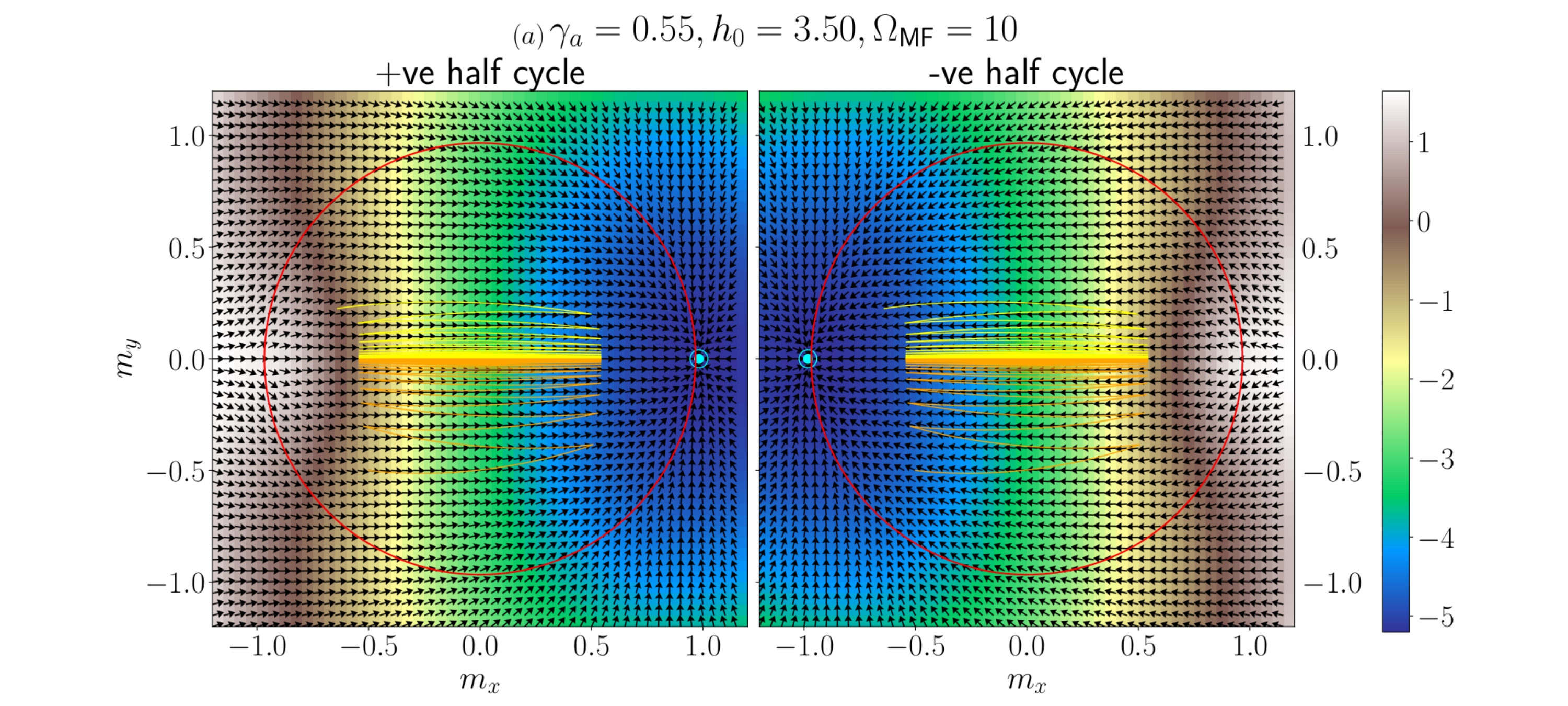}}
  \rotatebox{0}{\includegraphics[width=0.93\linewidth]{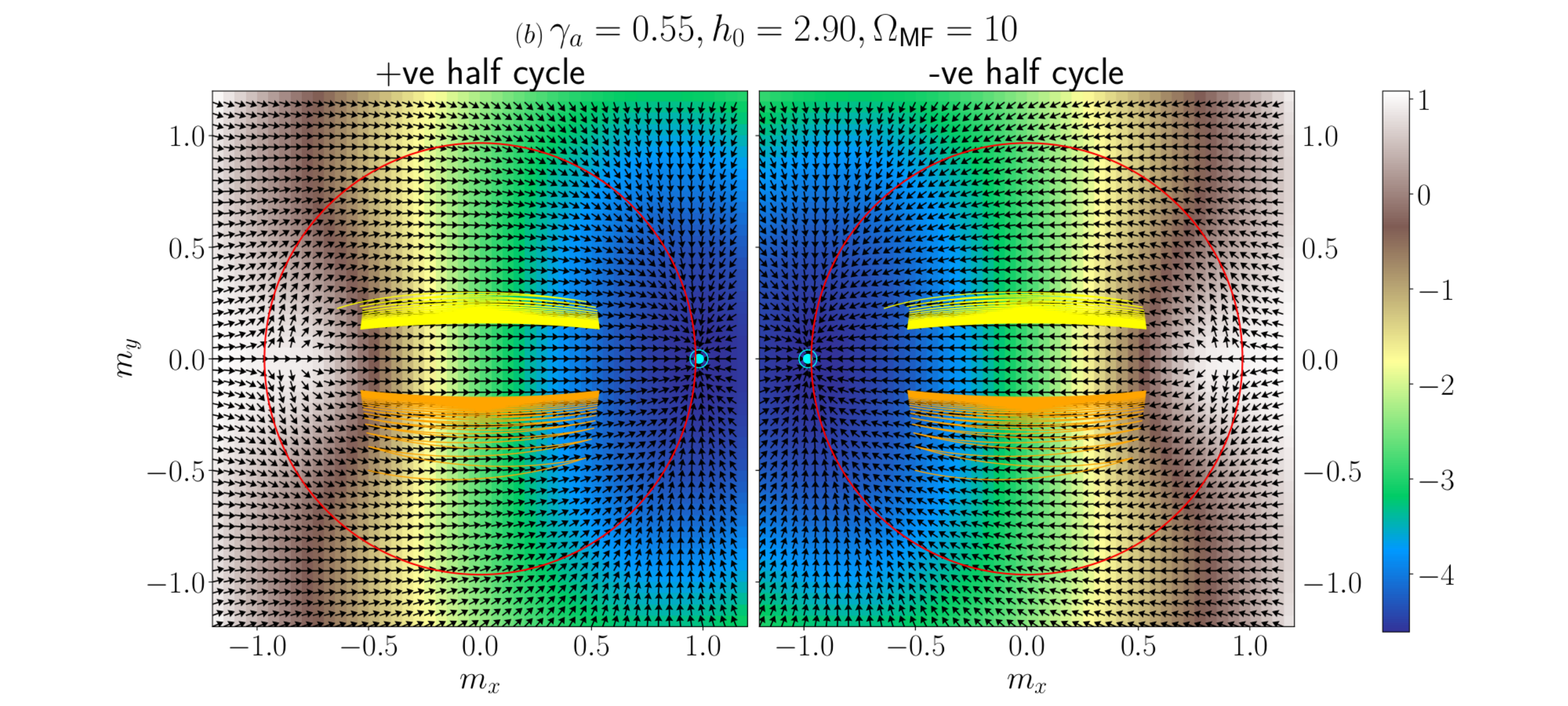}}
\caption{Free energy and fixed point structure corresponding to parameter choices $(a)\:\:\gamma_a=0.55,h_0=3.50$ and $(b)\:\:\gamma_a=0.55,h_0=2.90$ at a fixed frequency $\Omega_{\text{MF}}=10$. The background colors in the above plots indicate the local free energy density profile $f_{\text{MF}}(m_x,m_y)$ as a function of $m_x$ and $m_y$ (see Eq.~\eqref{eq:free-energy}). The left panel is for the positive half cycle of the drive, while the right panel shows the same for the negative half cycle. Since the external field is a square wave drive, the two panels describe the entire time dependence of $f_{\text{MF}}(m_x,m_y)$. The arrows indicate the direction of the local gradient of $f_{\text{MF}}(m_x,m_y)$, or effectively, the RHS of Eq.~\eqref{eq:XYdmdt}. In (a), orange and yellow lines represent two different trajectories that give the same Ising-SRO. In (b) the yellow and the orange trajectories represent two partner $\mathbb{Z}_2$ trajectories of the $XY$-SRO phase. In all plots we show the circle $\mathcal{C}_0$ (see Eq.~\eqref{eq:C0}), which represents the exact fixed points for $\gamma_a=h_0=0$. The fixed points for $h_0>0$ ($h_0<0)$ for the given parameter values are shown as blue dots with an open circle.}
\label{fig:fig4}
\end{figure*}

In Figs.~\ref{fig:fig3}(a)-\ref{fig:fig3}(d), we plot the absolute values of dynamic order parameters $|Q_x|$ and $|Q_y|$ with inverse system size $1/L$ for several parameter choices hosting $XY$-SBO phase as per Fig.~\ref{fig:fig1}. This shows us that, one of the two non-zero order parameters giving rise to the $XY$-SBO phase (either $Q_x$ or $Q_y$), vanishes systematically in the thermodynamic limit giving way to either Ising-SBO or $XY$-SRO when $L \rightarrow \infty$. In Figs.~\ref{fig:fig3}(e)-(f) we verify, using MC data, that the DPT is Ising like for $XY$-SRO to Ising-SRO transition (see Fig~\ref{fig:fig1}(a)-(c)). We compute this by performing the finite size scaling collapse of the susceptibility $\chi_y$ and the fourth order Binder cumulant $U_y^{(4)}$ as defined in Eq.~\eqref{eq:susc-defn}--\eqref{eq:U4-defn} respectively. Taking the critical exponents $\nu=1,\gamma=7/4$ and using scaling relations \eqref{eq:susc-scaling}--\eqref{eq:U4-scaling} we perform a finite size scaling collapse using the data for the different system sizes simultaneously to determine the only remaining free parameter $h^{\star}$. The critical field $h^{\star}$ is calculated by minimizing the reduced chi square per degree of freedom $\tilde\chi^2 =\frac{1}{N_d-M}\sum_{i=1}^{N_d}(y_i-f(x_i))^2/\sigma_i^2$ where $N_d$ equals the total number of data points, $M$ denotes the number of fitting parameters, $y_i=\chi_y L^{-\gamma/\nu}$ denotes the mean value of the i-th data point (LHS of Eq.~\ref{eq:susc-scaling}), $\sigma_i$ denotes the statistical error on the i-th data point, and $f(x_i)$ (RHS of Eq.~\ref{eq:susc-scaling}) denotes the fitting function which is taken to be a low-order polynomial (typically between 4th to 8th)~\cite{10.1063/1.3518900}. The error bar on $h^{*}$ is then obtained by repeating the $\tilde\chi^2$ minimization several times with Gaussian noise whose variance equals $\sigma^2_i$ at the i-th data point, to obtain several different values of $h^{*}$~\cite{10.1063/1.3518900}. The reliability of the procedure was checked both by (a) monitoring that $\tilde\chi^2 \sim O(1)$ as well as by (b) using the value of $h^{*}$ extracted from the data collapse of $\chi_y$ to verify whether a good data collapse is obtained for the $U_y^{(4)}$ data. The data collapse for $\chi_y$ and $U_y^{(4)}$ for $\gamma_a=0.55$ and $\Omega_{\mathrm{MC}}=2\pi/4$ using the aforementioned procedure is shown in Figs.~\ref{fig:fig3} (e) and (f), respectively. We obtain the critical value of magnetic field strength as $h^\star=2.245 \pm 0.001 $ for $\gamma_a=0,\Omega_{\text{MC}}=\pi/2$. We repeat the analysis at $\gamma_a=0.55,\Omega_{\text{\tiny{MC}}}=\pi/2$ which leads to the critical field strength $h^\star=2.030 \pm 0.001 $ corresponding to the $XY$-SRO to Ising-SRO transition in Fig~\ref{fig:fig1}(c). The absence of $XY$-SBO as $L \rightarrow \infty$ for the parameters used in Fig.~\ref{fig:fig1} (b) suggests that the DPT between Ising-SBO and $XY$-SRO is, in fact, weakly first-order for $\gamma_a=0.30$ and $\Omega_{\mathrm{MC}} = 2\pi/20$ since both $|Q_x|$ and $|Q_y|$ cannot be simultaneously tuned to be zero at the DPT by only varying $h_0$, unless the DPT is multicritical in nature. 

\begin{figure*}[!ht]
\includegraphics[width=0.75\textwidth]{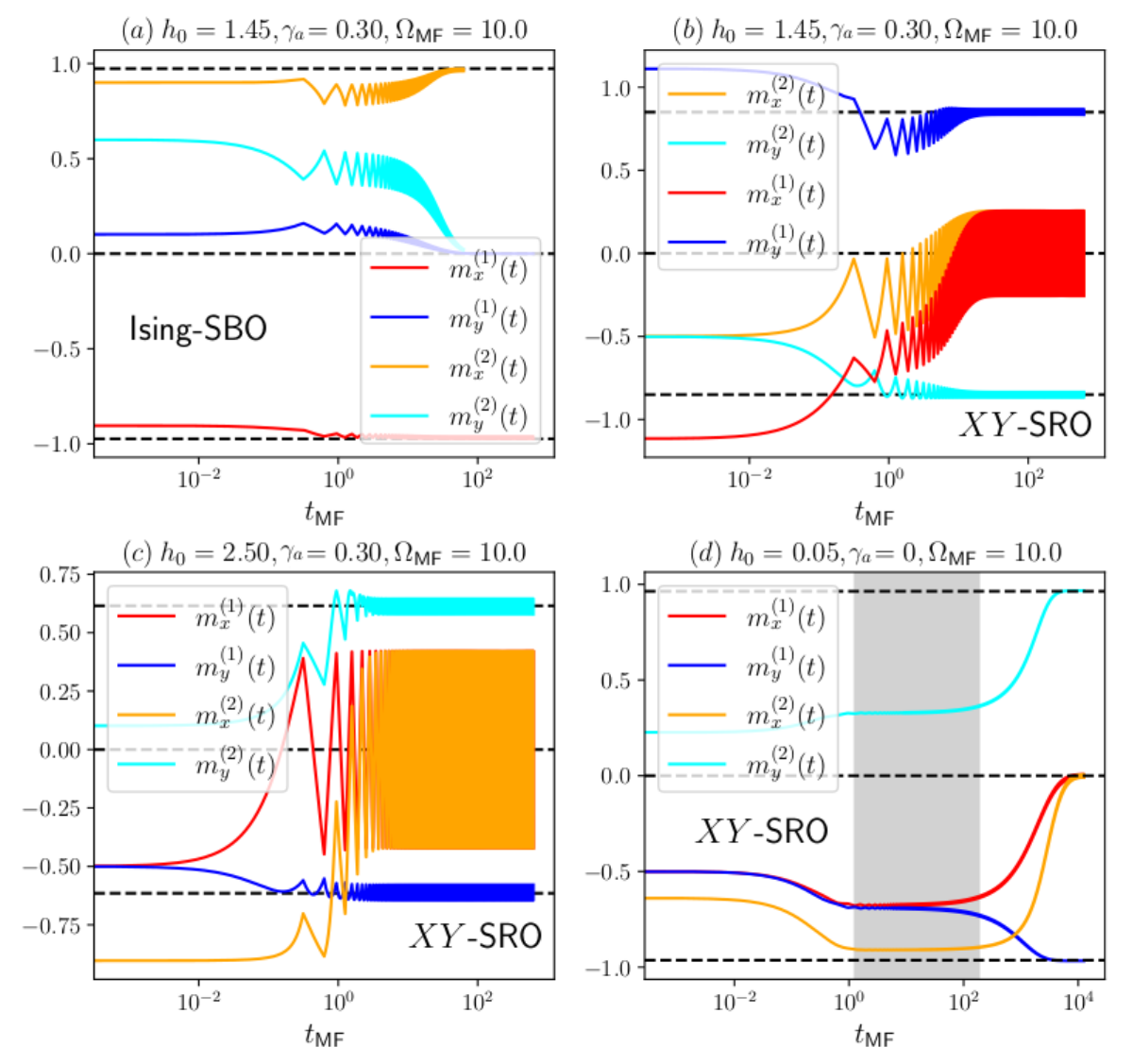}
\caption{MF trajectories for different parameter points obtained by solving Eq.~\eqref{eq:XYdmdt} starting from different initial conditions. In (a)-(d) the $x$ axis shows the continuous time as in Eq.~\eqref{eq:XYdmdt}, while $(m_x^{(r)}(t),m_y^{(r)}(t))$ for $r=1,2$ denotes the solutions of Eq.~\eqref{eq:XYdmdt} with two different initial condition. For (a) and (b) $\gamma_a=0.30,h_0=1.45,\Omega_{\text{MF}}=10$. Depending on the initial condition (see insets of Figs.~\ref{fig:fig1}(e), \ref{fig:fig1}(f), and text), the late time dynamics can be either Ising-SBO (a) or $XY$-SRO (b). Some initial conditions stabilize an Ising-SBO phase at late times while some of them stabilize $XY$-SRO phase. However, within each class of these trajectories (Ising-SBO or $XY$-SRO) the magnitude of the order parameter is the same. For (c) $\gamma_a=0.30,h_0=2.50,\Omega_{\text{MF}}=10$ the system does not show any substantial transient behavior, as the system quickly relaxes to $XY$-SRO phase. In (d) $\gamma_a=0,h_0=0.05,\Omega_{\text{MF}}=10$ we show the very slow approach to an $XY$-SRO trajectory, starting from an initial state. There exists a very large intermediate time window, where the dynamics displays transient $XY$-SBO, behaviour.}
\label{fig:fig5}
\end{figure*}

\begin{figure*}[!htbp]
    \centering\includegraphics[width=0.75\textwidth]{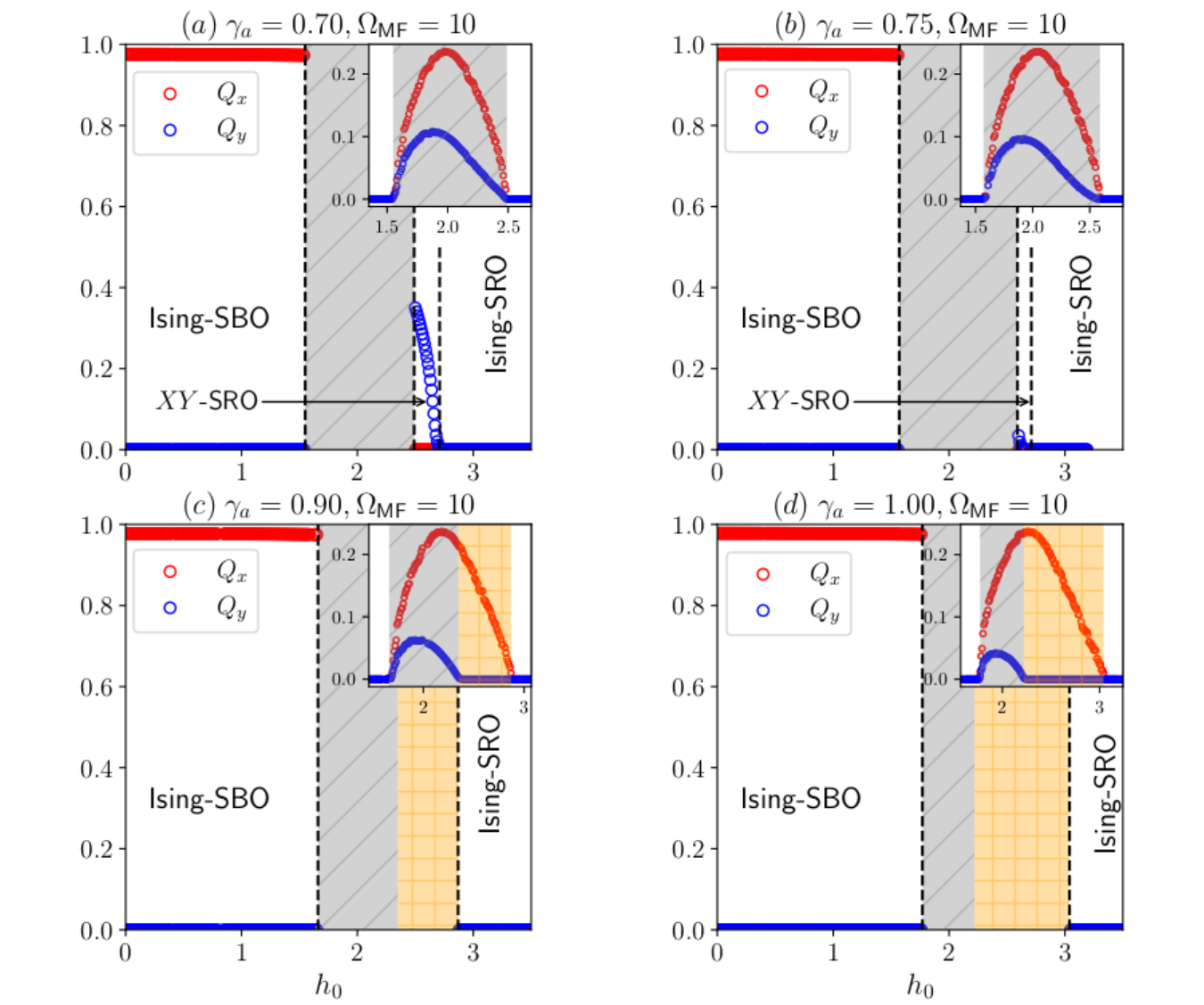}
    \caption{Some additional MF phase diagrams for $(a)\;\gamma_a=0.70$, $(b)\;\gamma_a=0.75$, $(c)\;\gamma_a=0.90$, $(d)\;\gamma_a=1.00$ at fixed $\Omega_{\text{MF}}=10$. The insets highlight the initial condition dependence by showing the variance of the order parameters $\text{var}(Q_x)$ (red circles) and $\text{var}(Q_y)$ (blue circles) for a uniform grid of 100 initial conditions with $m_{x,y} \in[-1,1]$ as a function of the magnetic field strength $h_0$ (same as Figs.~\ref{fig:fig1}(h) and \ref{fig:fig1}(i)). In the shaded gray regions, both $\text{var}(Q_x)$ and $\text{var}(Q_y)$  are non-zero while, in the shaded orange region indicates that only $\text{var}(Q_x)$ is non-zero. This illustrates that with increasing $\gamma_a$ the $XY$-SRO region shrinks continuously while the region of bistability between Ising-SBO and $XY$-SRO (shaded gray region) grows. After a cutoff value of $\gamma_a \gtrapprox 0.75 $ a new kind of bistability between Ising-SBO and Ising-SRO emerges (shaded orange region)}.
    \label{fig:fig6}
\end{figure*} 

\subsection{\label{sec:fixed-point-analysis} Mean-field analysis }

In contrast to the numerical MC simulations, coupled non-linear equations in $m_x(t)$ and $m_y(t)$ derived from a MF analysis offer a semi-analytical method (an alternative approach) to analyze the DPTs. In order to find the appropriate dynamical phase for a given parameter choice, the MF EOMs Eq.~\eqref{eq:XYdmdt} need to be evolved starting from some initial condition $(m_x^{(0)},m_y^{(0)})$. After an initial transient dynamics in time $(m_x(t),m_y(t))$ settles down to a periodic steady state with the same time period as the driven magnetic field. The system evolves to unique values for the magnitudes of the order parameters $|Q_x|$ and $|Q_y|$ from which the dynamical phases can be deduced. There is no guarantee that the order parameter magnitudes will be the same if we start from different initial conditions in the $(m_x,m_y)$ plane. The dynamical phase for a particular choice of couplings can be determined uniquely, and is called a stable dynamical phase, only if all the initial conditions (apart from certain measure zero initial conditions that will be discussed later) in the $(m_x,m_y)$ plane lead to the same values for $|Q_x|$ and $|Q_y|$ at late times. We encounter three stable dynamical phases from the MF EOMs, namely, Ising-SRO, Ising-SBO, and $XY$-SRO. From the numerical study of the evolution of generic initial conditions under the MF EOMs, we also find regions of {\it{bistability}} in the parameter space $(h_0,\gamma_a)$. In the region of bistability, the entire set of generic initial conditions can be divided into two subsets. Within each subset, all initial conditions flow to a unique periodic steady state with identical values for $(|Q_x|, |Q_y|)$. We find bistable regions between Ising-SBO and $XY$-SRO phases and between Ising-SBO and Ising-SRO phases, respectively. 

For Ising-SRO, Ising-SBO, as well as the bistable region between Ising-SRO and Ising-SBO, the MF EOMs lead to trajectories that concentrate along $m_y=0$ at late times while $XY$-like dynamical phases result in late-time trajectories that require $m_y \neq 0$. Two examples of the evolution of $(m_x(t),m_y(t))$ under the MF EOMs are shown in Fig.~\ref{fig:fig4} where Fig.~\ref{fig:fig4}(a) depicts Ising-SRO, which also shows the concentration of the trajectory along $m_y=0$ as time progresses, while Fig.~\ref{fig:fig4}(b) depicts $XY$-SRO, where the trajectory concentrates around $m_y \neq 0$ at late times to ensure $|Q_y| \neq 0$, with both panels showing two trajectories that lead to the same final values of $|Q_x|$ and $|Q_y|$. The concentration of trajectories along the line $m_y=0$ (Fig.~\ref{fig:fig4}(a)) or away from it (Fig.~\ref{fig:fig4}(b)) illustrates the non-trivial interplay between the switching of the fixed point(s) of the free energy $f_{\mathrm{MF}}(m_x,m_y)$ after every half cycle of the magnetic field and the nature of the gradient of $f_{\mathrm{MF}}(m_x,m_y)$. In both cases, the trajectories never converge to the fixed point of the free energy $f_{\mathrm{MF}}(m_x,m_y)$ during a half cycle since the system is driven well away from its adiabatic limit of $\Omega_{\mathrm{MF}} \rightarrow 0$.

The nature of the dynamical phases and the bistable regions from the MF EOMs can be anticipated by concentrating on trajectories with the initial condition of the form $(m_x, \delta m_y)$ where $|\delta m_y| \ll 1$. Let us first restrict to initial conditions of the form $(m_x,0)$. It may be deduced from the form of Eq.~\eqref{eq:XYdmdt} and explicit computations that the fixed points of $-Jqm_\alpha +G_\alpha(m_x,m_y)=0$, where $\alpha=x,y$, lie along the $m_y=0$ axis. Using $G_y(m_x,m_y=0)=0$ and using certain identities for the modified Bessel functions, one can show that the fixed point(s) in $m_x$ (with $m_y$ being zero) are 
given by the solution(s) of the following equation
\begin{equation} \label{eq:fixed_point_zero_my_simplified}
    m_x = \frac{\sum\limits_{\mu=-\infty}^{\infty} \mathcal{I}_{\mu}(\beta\gamma_a/2)\mathcal{I}_{2\mu+1}(\beta h_{\text{eff}}^x) }{\sum\limits_{\mu=-\infty}^{\infty} \mathcal{I}_{\mu}(\beta\gamma_a/2)\mathcal{I}_{2\mu}(\beta h_{\text{eff}}^x) }.
 \end{equation}
Solving Eq.~\eqref{eq:fixed_point_zero_my_simplified} numerically for $h<h^*(\gamma_a,\beta)$ yields three fixed points $-$ one attractive, one repulsive, and one saddle point (attractive along $m_x$ but repulsive along $m_y$). But, for $h>h^*(\gamma_a,\beta)$, there is only one attractive fixed point. Solving MF equations Eq.~\eqref{eq:XYdmdt}(a) and \eqref{eq:XYdmdt}(b) for a trajectory with $m_y=0$ then gives $m_y=0$ for all times while $m_x$ satisfies an equation similar in structure to that of an Ising MF EOM.  This yields an Ising-SBO (Ising-SRO) for $h<h^*(\gamma_a,\beta)$ ($h>h^*(\gamma_a,\beta)$) in the adiabatic limit of low drive frequencies. For moderate drive frequencies, the Ising-SBO and the Ising-SRO phases can be separated by a bistable region between Ising-SBO and Ising-SRO as a function of $h$ where a class of trajectories converge to an Ising-SBO while the rest to an Ising-SRO. 

In the analysis of the previous paragraph, we restricted ourselves to initial conditions with $m_y=0$ which are measure zero in a 2d phase phase $(m_x,m_y)$. One way to remedy this is to consider a class of initial states with $(m_x,\delta m_y)$ where $|\delta m_y| \ll 1$ is a small number. Using such initial conditions we can propagate the MF EOMs. Concentrating on the evolution of $|\delta m_y(t)|$ as a function of time, we see three kinds of behaviors from our MF EOMs: (a) for Ising-SBO/ Ising-SRO phases  or for bistable regions between Ising SBO and Ising-SRO, the perturbation $\delta m_y$ decreases exponentially as a function of time while for $XY$-SRO phase, the perturbation increases in time and eventually saturates to a non-zero value. The behavior of $|\delta m_y(t)|$ is most interesting for the bistable region between Ising-SBO and $XY$-SRO where for Ising-SBO trajectories, $|\delta m_y(t)|$ decreases exponentially in time while for $XY$-SBO trajectories, $|\delta m_y(t)|$ grows in time and saturates.

Going beyond initial conditions with small $\delta m_y$, further examples of MF trajectories starting from some specific initial conditions are shown in Fig.~\ref{fig:fig5} for different values of $\gamma_a$ and $h_0$. While both Figs.~\ref{fig:fig5} (a), (b) show the behavior of MF trajectories at a parameter value deep in the bistable region,  Fig.~\ref{fig:fig5} (c) is for a parameter value deep in the $XY$-SRO region while Fig.~\ref{fig:fig5} (d) is for $\gamma_a=0$ and a small $h_0$ that leads to a pre-thermal $XY$-SBO behavior before the trajectories eventually converge to $XY$-SRO.  Fig.~\ref{fig:fig5} (c) shows that the transients die out at $\mathcal{O}(1)$ times when measured in units of $J^{-1}$ and different trajectories settle close to the same final steady state value when the parameter values are away from any DPTs or regions of bistability. On the other hand, the trajectories in Fig.~\ref{fig:fig5}(a) (Fig.~\ref{fig:fig5}(b)) display Ising-SBO ($XY$-SRO) behavior with two different initial conditions displaying identical magnitudes of the respective order parameters, $|Q_x|$ and $|Q_y|$ at late times $t_{\mathrm{MF}} \gg 1$. Fig.~\ref{fig:fig5}(d) shows a long prethermal $XY$-SBO behavior in time (marked in grey) where the magnitude of the order parameters $|Q_x|, |Q_y|$ depend on the initial condition, which finally gives way to $XY$-SRO for $t_{\mathrm{MF}} \gg 1$ where $|Q_x|, |Q_y|$ are independent of the initial conditions. The data displayed in Figs.~\ref{fig:fig5}(a), (b), (d) clearly show that $XY$-SBO, which requires both $Q_x$ and $Q_y$ to be non-zero, emerges only as a transient dynamical phase within the MF EOMs. We refer the reader to Fig.~\ref{fig:fig2}(b) for the MF phase diagram in the $(h_0,\gamma_a)$ plane for a fixed $\Omega_{\mathrm{MF}}=10$ and $\beta=4$. 

In Figs.~\ref{fig:fig6} (a)-(d), we show more numerical details for the region of bistability predicted from the MF EOMs for four different values of $\gamma_a$ as $h_0$ is varied. Just like Figs.~\ref{fig:fig2}(e)-(f) (insets), the insets of Figs.~\ref{fig:fig6} (a)-(d) show the variance of both $|Q_x|$ and $|Q_y|$ calculated using $100$ initial conditions on a uniform grid formed by $m_x \in [-1,1]$ and $m_y \in [-1,1]$ at each of the parameter values to quantify bistability. For $\gamma_a=0.70$ (Fig.~\ref{fig:fig6}(a)) and $\gamma_a=0.75$ (Fig.~\ref{fig:fig6}(b)), we find that increasing $h_0$ leads to Ising-SBO, followed by a region of bistability (indicated by the shaded grey regions), $XY$-SRO and finally Ising-SRO. A generic initial condition in the bistable regions of Fig.~\ref{fig:fig6}(a) and (b) can be divided into two subsets, where initial conditions in each subset flow to either Ising-SBO or $XY$-SRO characterized by identical values of $(|Q_x|, |Q_y|)$.  The width of the $XY$-SRO regions decreases with increasing $\gamma_a$ and for $\gamma_a \gtrapprox 0.75$, the $XY$-SRO region disappears altogether (Figs.~\ref{fig:fig6} (c)-(d)), with the bistability regions being flanked by Ising-SBO and Ising-SRO on both sides. The nature of the bistable region is, however, more interesting as the corresponding insets show. The grey shaded regions are bistable regions where the two subsets of generic initial conditions flow to either Ising-SBO or $XY$-SRO as before, while in the orange shaded regions, the two subsets flow to either Ising-SBO or Ising-SRO. At this point, it is useful to stress that while the bistability between Ising-SBO and Ising-SRO can also be understood from an effective one-dimensional EOM, the bistability between Ising-SBO and $XY$-SRO crucially relies on the two-dimensional nature of the phase space for the MF EOMs.  

We finally discuss the prethermal $XY$-SBO obtained for small $h_0$ when $\gamma_a=0$ (Fig.~\ref{fig:fig1}(d) and Fig.~\ref{fig:fig5}(d)). For $\gamma_a=0$, the MF free energy is given by 
\begin{equation}
        f_{\text{MF}}(m_x,m_y) = \frac{Jq\vec{m}^2}{2} - \frac{1}{\beta} \ln \Big( \mathcal{I}_0(\beta h_{\text{eff}}) \Big).
        \label{eq:free-energy-isotropic}
\end{equation}
By definition, the \textit{fixed points} of the free energy correspond to points on the $m_x-m_y$ plane where the gradients of the MF free energy $f_{\text{MF}}(m_x,m_y)$ vanish \begin{subequations}
\begin{align}
    \label{eq:fixed-points-zero-anisotropy-mx}
    \frac{\partial f_{\text{MF}}}{\partial m_x} &= Jq\left( m_x - \frac{\mathcal{I}_1(\beta h_{\text{eff}})}{\mathcal{I}_0(\beta h_{\text{eff}})} \frac{Jqm_x+h_{\text{ext}}}{h_{\text{eff}}} \right) = 0, \\
    \frac{\partial f_{\text{MF}}}{\partial m_y} &= Jq\left( m_y - \frac{\mathcal{I}_1(\beta h_{\text{eff}})}{\mathcal{I}_0(\beta h_{\text{eff}})} \frac{Jqm_y}{h_{\text{eff}}}\right) = 0. 
    \label{eq:fixed-points-zero-anisotropy-my}
\end{align}
\end{subequations} To solve Eq.~\eqref{eq:fixed-points-zero-anisotropy-my} we implement the following two conditions $(i) \; \displaystyle m_y=0$ and $(ii) \; \displaystyle \mathcal{I}_1(\beta h_{\text{eff}})\; Jq/\mathcal{I}_0(\beta h_{\text{eff}})h_{\text{eff}}=1$. The second condition cannot be satisfied simultaneously with Eq.~\eqref{eq:fixed-points-zero-anisotropy-mx}, unless $h_{\text{ext}}=0$. This corresponds to the standard equilibrium $2d$-$XY$ model, which we shall address shortly. Putting $m_y=0$ in Eq.~\eqref{eq:fixed-points-zero-anisotropy-mx} we find $m_x = \mathcal{I}_1(\beta h_{\text{eff}})/\mathcal{I}_0(\beta h_{\text{eff}})$. This equation can be solved numerically to find the location of the fixed points, if any.  

Next, we shall discuss the case of exactly zero external field (standard $2d-XY$ model). In this case, the fixed points are solutions of the following transcendental equation \begin{equation}
    \frac{\mathcal{I}_1(\beta h_{\text{eff}
    })}{\mathcal{I}_0(\beta h_{\text{eff}})}\frac{Jq}{h_{\text{eff}}}=1,
\end{equation}
which can be rewritten as \begin{equation}
    \frac{\mathcal{I}_1(u)}{\mathcal{I}_0(u)}\frac{1}{u} = \frac{1}{\beta J q},
    \label{eq:circle-u}
\end{equation}
where we have defined $u=\beta h_{\text{eff}}$. The solutions of Eq.~\eqref{eq:circle-u} can be found numerically. Assuming such a solution exists at $u=u^{\ast}$, we can immediately conclude that all the fixed points $(m_x^{\ast},m_y^{\ast})$ lie on the circle $\mathcal{C}_0$ defined by \begin{equation}
    {m_x^{\ast}}^2 + {m_y^{\ast}}^2 = \left( \frac{u^{\ast}}{\beta J q} \right)^2.
    \label{eq:C0}
\end{equation} Numerically solving Eq.~\eqref{eq:circle-u}, for $\beta=4,J=1,q=4$ we find $u^{\ast}/\beta J q = 0.96712 \pm 0.00001 $

The MF EOMs predict a long-lived prethermal $XY$-SBO for small $\gamma_a$ and $h_0$ that eventually melts into an $XY$-SRO. The timescale of this transient regime, $\tau_{\text{pth}}$, is obtained by fitting $|m_x(t)| \sim \exp(-t/\tau_{\text{pth}})$ at late times and is shown in the inset of Fig.~\ref{fig:fig1}(d) for $\gamma_a=0$. While the magnitude of $m_x(t)$ at time $t$ does seem to be dependent on the initial condition, $\tau_{\text{pth}}$ is fairly insensitive to it giving a precise meaning to the prethermal timescale. This prethermal regime can be loosely understood as follows. For $\gamma_a=0$ and $h_0=0$, Eq.~\eqref{eq:C0} yields an infinite number of fixed points instead of either one or three, where all these fixed points $(m_x^*, m_y^*)$ lie on a circle $\mathcal{C}_0$ of a fixed radius. For small $\gamma_a$ and $h_0$, a generic initial condition first yields a ``fast'' motion towards this circle $\mathcal{C}_0$ followed by a ``slow'' transient motion along this circle (which defines the prethermal $XY$-SBO) before the system eventually settles into $XY$-SRO.

It is worth mentioning that in certain regimes of the parameters $h_0,\gamma_a$, and $\Omega$ such as when one of the couplings dominate the other two (i.e., $h_0 \gg \gamma_a, \Omega$ or $h_0 \ll \gamma_a, \Omega$), one can intuitively deduce the late time steady state using simple arguments. By understanding the magnetization reversal time $\tau_{\text{flip}}$ behaviour in these regimes and comparing it with the half period of external drive $\pi/\Omega$, one can estimate what type of Ising phases would be preferred by the system. If $\tau_{\text{flip}} \gg \pi/\Omega$, which is expected for $h_0 \ll \gamma_a, \Omega$, then Ising SBO phase is the late time steady state, while in cases where $\tau_{\text{flip}} \ll \pi/\Omega$, which is expected for $h_0 \gg \gamma_a, \Omega$, we observe Ising SRO trajectories. We have verified that this expectation is borne out directly from the MF EOM approach as well. However this kind of simple argument can only ever predict Ising trajectories even in cases where the true steady state is $XY$ like. Since, we are using only one timescale comparison, we can only distinguish between two different dynamical phases and not all four of them. Also, in the inherent way we define magnetization reversal time $\tau_{\text{flip}}$, we assume a micromotion which is consistent with one of the two Ising phases while completely disregarding any $XY$ like trajectory. However, when this simple argument works, it can be extremely accurate. For example, in \cite{BAEZ201015}, such arguments were used to locate the critical couplings for a kinetic Ising model. 

\section{\label{sec:conc}Conclusion} We have investigated the non-trivial dynamical phases and the phase transitions that can arise in a non-equilibrium classical $2d$ an-$XY$ model driven by an external magnetic field. Utilizing the Glauber algorithm in a MC simulation implemented within a CPU + GPU heterogeneous computing paradigm, we identify the occurence of four dynamical phases and three phase transitions. Of the four DPT phases, three are thermodynamically stable (Ising-SBO, Ising-SRO, and $XY$-SRO). However, one of the phases $XY$-SBO vanishes in the thermodynamic limit (based on finite size extrapolation). The MC simulations suggest that all the phase transitions belong to the Ising universality class or are first-order in nature (based on FSS). There is also a hint that a tricritical point can exist in the anisotropy-magnetic field phase diagram with three different dynamical phases in its neighborhood (Fig.~\ref{fig:fig2}(a)). Future works can explore whether a non-Ising critical behavior is obtained here using finite size scaling analysis.

We supplement our MC results with a non-linear coupled MF equation analysis of the driven $2d$ an-$XY$ model. Mean-field analysis supports the existence of all the phases. The phase diagram qualitatively agrees with the MC results. As in the MC simulation, the $XY$-SBO phase displays a non-trivial physical behavior. Within MF it arises in region of the free energy flow diagram where it is susceptible to disintegrating into the Ising-SBO or the $XY$-SRO phase. We find two sets of distinct initial conditions for which this flow pattern happens. Thus, the $XY$-SBO phase generates a non-trivial $\mathbb{Z}_2$ bifurcation of the MF equation fixed point structure.

Finally, we note that in an externally driven classical spin model, dynamical phase transitions may be present due to the competition of the time period of the external field and the system's own relaxation timescales. In the case of the kinetic Ising model, this is due to a single- or multi- droplet mechanism~\cite{SidesPhysRevLett.81.834,PleimlingPhysRevE.87.032145}. However, in the case of spins with a continuous rotational degree of freedom such as in Eq.~\eqref{eq:hamXY}, the droplet picture breaks down. It remains to be understood what is the underlying mechanism contributing to the dynamical phases and the phase transitions that are observed.

\begin{acknowledgements}
MP and AS acknowledge the support of the IACS High Performance Computing services for providing computational resources contributing to the results presented in this publication. TD acknowledges Augusta University High Performance Computing Services (AUHPCS) for providing computational resources contributing to the results presented in this publication. WDB and TD acknowledges NSF-MRI\#0922362 and Cottrell College Science Award. TD acknowledges the hospitality of KITP at UC-Santa Barbara. A part of this research was completed at KITP and was supported in part by the National Science Foundation under Grant No. NSF PHY-1748958. AS thanks Jayanta K. Bhattacharjee for a useful discussion. TD thanks Mark Novotny for useful conversations.
\end{acknowledgements}
    
\bibliography{references}

\newpage
\begin{appendix}
\section{Mean-field theory}{\label{appendix-MFT}}\label{appendix a}

\subsection{Calculation of $\mathcal{Z}_{\text{MF}}$}{\label{appendix-derivation-ZMF}} Using standard identities related to the modified Bessel functions (see \cite{abramowitz+stegun}), we can write the exponential factors within the integral on the RHS of Eq.~\eqref{eq:MF-partition} as \begin{equation}
    \label{eq:ZMF-derivation-exp-heff}
    \exp\Big( \beta h_{\text{eff}}\cos(\theta-\phi) \Big) = \sum_{\nu=-\infty}^{+\infty} \mathcal{I}_{\nu}(\beta h_{\text{eff}}) e^{i\nu(\theta-\phi)},
\end{equation}
\begin{equation}
    \label{eq:ZMF-derivation-exp-gamma}
    \exp\Big( \beta\gamma_a \cos^2\theta \Big) = \sum_{\mu=-\infty}^{+\infty} \mathcal{I}_{\mu}(\beta\gamma_a/2) e^{2i\mu\theta} \exp(\beta\gamma_a/2).
\end{equation}
Thus, we have 
\begin{widetext}
\begin{equation}
    \begin{split}
            \mathcal{Z}_{\text{MF}} \propto \exp\Big(\frac{\beta JNq\vec{m}^2}{2}\Big) & \Bigg[ \sum_{\nu,\mu=-\infty}^{\infty} \mathcal{I}_{\nu}(\beta h_{\text{eff}})\: \mathcal{I}_{\mu}(\beta\gamma_a/2) 
           \int_0^{2\pi} d\theta\; e^{i\theta(\nu+2\mu)-i\nu\phi} \Bigg]^N,
    \end{split}
\end{equation}
\end{widetext}
which leads to Eq.~\eqref{eq:MF-partition-simplified}.

\subsection{Magnetization Dynamics} The phenomenological EOMs governing the dynamics of the system within a MF approximation are given by 
\begin{equation}
\label{eq:appendix-magnetization-dynamics-MFT-mx}
\Gamma_\alpha \frac{dm_\alpha}{dt}=-\frac{\partial f_{\text{MF}}}{\partial m_\alpha} = -Jqm_\alpha + G_{\alpha}(m_x,m_y), 
\end{equation}
where $\alpha = x,y$ and $f_{\text{MF}}(m_x,m_y)$ is given by Eq.~\eqref{eq:free-energy} and $G_x(m_x,m_y)$ and $G_y(m_x,m_y)$, defined in Eq.~\eqref{eq:appendix-GxGy}, come from the differentiation of the terms involving the modified Bessel functions. Explicit forms for $G_x(m_x,m_y)$ and $G_y(m_x,m_y)$ are given in Eq.~\eqref{eq:appendix-GxGy}.  For computing the derivative we used standard recursion relations (see \cite{abramowitz+stegun}). The resulting EOMs given in Eq.~\eqref{eq:appendix-magnetization-dynamics-MFT-mx} are then solved numerically using standard ODE solvers, taking the time dependence of the external field as a square wave of amplitude $h_0$ and frequency $\Omega_{\text{MF}}$. Formally $G_x,G_y$ are both ratios of infinite sums of modified Bessel functions of all orders. However, for the purpose of numerically solving Eq.~\eqref{eq:appendix-magnetization-dynamics-MFT-mx} we have evaluated $G_x,G_y$ by truncating the said infinite sums in the numerator and denominator, independently within a convergence precision of $10^{-8}$.

\begin{widetext}
\begin{subequations}
\label{eq:appendix-GxGy}
\begin{align}
    G_x(m_x,m_y) = \frac{\sum\limits_{\mu=-\infty}^{\infty} \mathcal{I}_{\mu}(\beta\gamma_a/2)\Big[ \frac{2\mu}{\beta h_{\text{eff}}}\mathcal{I}_{2\mu}(\beta h_{\text{eff}}) + \mathcal{I}_{2\mu+1}(\beta h_{\text{eff}}) \Big]\frac{\beta Jqh^x_{\text{eff}}}{h_{\text{eff}}}\cos(2\mu\phi) +2\mu \mathcal{I}_{\mu}(\beta\gamma_a/2)\mathcal{I}_{2\mu}(\beta h_{\text{eff}})\sin(2\mu\phi)\frac{Jqh_{\text{eff}}^y}{h_{\text{eff}}^2} }{\beta\sum\limits_{\mu=-\infty}^{\infty} \mathcal{I}_{\mu}(\beta\gamma_a/2)\mathcal{I}_{2\mu}(\beta h_{\text{eff}}) \cos(2\mu\phi) } \\
    G_y(m_x,m_y) = \frac{\sum\limits_{\mu=-\infty}^{\infty} \mathcal{I}_{\mu}(\beta\gamma_a/2)\Big[ \frac{2\mu}{\beta h_{\text{eff}}}\mathcal{I}_{2\mu}(\beta h_{\text{eff}}) + \mathcal{I}_{2\mu+1}(\beta h_{\text{eff}}) \Big]\frac{\beta Jq h_{\text{eff}}^y}{h_{\text{eff}}}\cos(2\mu\phi) - 2\mu \mathcal{I}_{\mu}(\beta\gamma_a/2)\mathcal{I}_{2\mu}(\beta h_{\text{eff}})\sin(2\mu\phi)\frac{Jqh_{\text{eff}}^x}{h_{\text{eff}}^2} }{\beta\sum\limits_{\mu=-\infty}^{\infty} \mathcal{I}_{\mu}(\beta\gamma_a/2)\mathcal{I}_{2\mu}(\beta h_{\text{eff}}) \cos(2\mu\phi) }
\end{align}
\end{subequations}
\end{widetext}

\subsection{Ising Limit $(\gamma_a\rightarrow\infty)$}{\label{appendix-ising-limit}}
The model Hamiltonian in Eq.~\eqref{eq:hamXY} corresponds to that of the an-$XY$ model. In the limit $\gamma_a\rightarrow\infty$ however, this should just become the Ising model. This happens as any non-zero $y$ component of field would be infinitely expensive energetically and hence the system only points in either +$x$ or -$x$ direction. The standard free energy of the Ising model should then be recovered from $\gamma_a\rightarrow\infty$ limit of Eq.~\eqref{eq:free-energy}. Here we show that this is indeed the case. First we note that as the only allowed equilibrium value of $m_y$ is zero, we can set $\phi=0$ and hence $h_{\text{eff}} = h_{\text{eff}}^x$. So in equilibrium,  when $dm_x/dt=0=-Jqm_x+G_x(m_x,0)$, the equilibrium value of $m_x$ is found from the solution of the following transcendental equation
    
\begin{equation}
\label{eq:ising-limit-mx}
             Jqm_x=\frac{1}{\beta}\frac{\mathcal{I}_0(\beta\gamma_a/2)\sum\limits_{\mu=-\infty}^{\infty}\frac{\mathcal{I}_{\mu}(\beta\gamma_a/2)}{\mathcal{I}_{0}(\beta\gamma_a/2)}\frac{\beta Jq h_{\text{eff}}^x}{h_{\text{eff}}}\mathcal{I}_{2\mu+1}(\beta h_{\text{eff}})}{\mathcal{I}_0(\beta\gamma_a/2)\Bigg[ \mathcal{I}_0(\beta h_{\text{eff}}) + \sum\limits_{\mu=-\infty}^{\infty} \mathcal{I}_{2\mu}(\beta h_{\text{eff}}) \frac{\mathcal{I}_{\mu}(\beta\gamma_a/2)}{\mathcal{I}_{0}(\beta\gamma_a/2)}\Bigg]}.
\end{equation} Now $\mathcal{I}_{\mu}(\beta \gamma_a/2)/\mathcal{I}_{0}(\beta \gamma_a/2) \rightarrow 1$ as $\gamma_a\rightarrow\infty$. After simplifications, Eq.~\eqref{eq:ising-limit-mx} reduces to \begin{equation}
\begin{split}
    m_x & = \frac{\sum\limits_{\mu=-\infty}^{\infty}\mathcal{I}_{2\mu+1}(\beta h_{\text{eff}})}{ \mathcal{I}_0(\beta h_{\text{eff}}) + \sum\limits_{\mu=-\infty}^{\infty} \mathcal{I}_{2\mu}(\beta h_{\text{eff}})}\\
    &= \frac{2\sum\limits_{\mu=0}^{\infty}{\mathcal{I}_{2\mu+1}(\beta h_{\text{eff}}) }}{\mathcal{I}_0(\beta h_{\text{eff}})+2\sum\limits_{\mu=1}^{\infty}{\mathcal{I}_{2\mu}(\beta h_{\text{eff}})}}= \tanh(\beta h_{\text{eff}}),
\end{split}
\end{equation} which is the well known transcendental equation governing the spontaneous magnetization of the Ising model, in the MF approximation. 

\end{appendix}

\end{document}